%% file: comparison_data_length.tex
\RequirePackage{fix-cm}
\documentclass[smallextended]{svjour3}       
\smartqed  
\usepackage{graphicx}
 \usepackage{amsmath}
 \usepackage{subfigure}
 \usepackage[authoryear]{natbib}
 \usepackage{xcolor}
 \usepackage{bm}
 \usepackage{url}
 \usepackage{verbatim} 
 \usepackage{float} 
\usepackage{ulem}

\begin{document}

\title{A comparative study of the robustness of frequency--domain connectivity measures to finite data length.
}

\author{Sara Sommariva$^{1,2}$ \and Alberto Sorrentino$^{1}$ \and Michele Piana$^{1}$ \and Vittorio Pizzella$^{3, 4}$ \and Laura Marzetti$^{3, 4}$ \\
\vspace{0.3cm} \\
\small
$^1$ Dipartimento di Matematica, Universit\`a degli studi di Genova \\
$^2$  Department of Neuroscience and Biomedical Engineering, Aalto University\\
$^3$ Department of Neuroscience, Imaging and Clinical Sciences, “G. d'Annunzio” University of Chieti-Pescara\\
$^4$Institute for Advanced Biomedical Technologies, “G. d'Annunzio” University of Chieti-Pescara}


\institute{}


\maketitle

\begin{abstract}
In this work we use numerical simulation to investigate how the temporal length of the data affects the reliability of the estimates of brain connectivity from EEG time--series. We assume that the neural sources follow a stable MultiVariate AutoRegressive model, and consider three connectivity metrics: Imaginary part of Coherency (IC), generalized Partial Directed Coherence (gPDC) and frequency--domain Granger Causality (fGC). In order to assess the statistical significance of the estimated values, we use the surrogate data test by generating phase--randomized and autoregressive surrogate data.
We first consider the ideal case where we know the source time courses exactly. Here we show how, expectedly, even exact knowledge of the source time courses is not sufficient to provide reliable estimates of the connectivity when the number of samples gets small; however, while gPDC and fGC tend to provide a larger number of false positives, the IC becomes less sensitive to the presence of connectivity.
Then we proceed with more realistic simulations, where the source time courses are estimated using eLORETA, and the EEG signal is affected by biological noise of increasing intensity. Using the ideal case as a reference, we show that the impact of biological noise on IC estimates is qualitatively different from the impact on gPDC and fGC.
\end{abstract}

\input{introduction.tex}

\input{methods.tex}

\input{results.tex}

\input{discussion.tex}

\input{conclusions.tex}

\begin{acknowledgements}
The authors thank the anonymous reviewers for their valuable comments and suggestions.\\
L.M. and V.P. have been supported in part by the grant \textit{Functional connectivity and 
neuroplasticity in physiological and pathological aging}, PRIN 2010–2011 n. $2010SH7H3F\_006$, and by the grant \textit{Breaking the 
Nonuniqueness Barrier in Electromagnetic Neuroimaging} (BREAKBEN), H2020-FETOPEN-2014-2015/H2020-FETOPEN-2014-2015-RIA, Project reference: 686865. 

\end{acknowledgements}

\bibliographystyle{spbasic}      

\bibliography{biblio_connectivity} 

\end{document}

%% file: introduction.tex
\section{Introduction}
\label{intro}

The idea that the synergic cooperation of several regions is required for the brain to be able to instantiate  specific functions and behaviors has, in the recent years, become central to neuroscience. Largely interconnected brain networks have indeed been reported to act as building blocks for the dynamic segregation and integration of brain areas during task execution or at rest in a wide range of spatial and temporal scales \citep{engel_13}. Understanding brain connectivity, both at structural and functional level, is thus a prerequisite for understanding brain functioning as well as its alterations \citep{stam_10}. 
In this framework, magnetoencephalography (MEG) and electroencephalography (EEG) have significantly contributed to unravel the functional wiring of the brain by putting emphasis on its temporal aspects, and looking for mechanisms of oscillatory coupling (in the range of about 1 to 100 Hz) as well as for coupled slower aperiodic fluctuations of brain activity \citep{engel_13}. To robustly measure functional coupling in task related or ongoing brain activity, a lot of effort has thus been made in the recent years in the development of methods for MEG and EEG connectivity with the aim of capturing one aspect or the other \citep{marzetti_08, stam_12, hillebrand_12, ewald_12, marzetti_13, chella_14, brookes_11, brookes_11b, brookes_12, depasquale_10, depasquale_12}. Several time--domain or frequency--domain metrics have thus been designed to capture the statistical dependencies or the causal relationships among EEG and MEG time series \citep{pereda_05, sakkalis_11, vanDiessen_15}. Despite the flourishing of such methods, still several issues affect the robustness of the estimation \citep{Schoeffelen_2009, basa14} which require systematic investigation and  comparison between the different metrics in order to assess their performance in terms of statistical robustness. To date, only few studies have faced this issue.\\ 
\noindent
For instance, \cite{wendling_09} and \cite{silfverhuth_12} study the robustness of some connectivity measures with respect to the relative strength of the interaction and of the innovation term in the underlying MultiVariate AutoRegressive (MVAR) model;
\cite{brookes_11} compare different connectivity measures in simulations and on resting state MEG data to test for robustness and similarities with respect to fMRI resting state networks; \cite{chella_14} introduce a novel third--order spectral connectivity measure and compare  its statistical robustness to traditional frequency--domain measures by means of simulated and experimental EEG data; more recently \cite{hietal16} use synthetic MEG data to compare the best regularization parameter required in order to estimate either the auto--spectral density function or the coherence function from the neural sources reconstructed by means of the Minimum Norm Estimate. Even fewer studies explicitly investigate the robustness of connectivity estimates with respect to the length of the data; \cite{astolfi_07} use synthetic data to compare three connectivity measures all based on an MVAR model; they exploit knowledge of the theoretical values of the considered connectivity measures to compute relative estimation errors; among other results, they quantify the minimum number of (possibly non--consecutive) time--points that are necessary to obtain good estimates of the connectivity patterns. \cite{bonita_14} compare the robustness to epoch length of four time--domain connectivity measures, using experimental resting state data: the criterion of comparison is their ability to discriminate between eyes open and eyes closed conditions. Among other results, they show on a toy example how considering too long data can result in a wrong estimation of the connectivity measures, due to the changes in the underlying connectivity pattern. Similarly, \cite{fretal16} use resting--state EEG recordings to investigate the impact of the epoch length on two different measures of functional connectivity. They show that this impact can be reduced by performing connectivity analysis in the source space and using a proper metric to characterize the topology of the estimated network. \cite{wang_14} present a Matlab toolbox that allows to compare the performances of many connectivity measures, using synthetic data generated with several different models, and use this toolbox to find the minimum data length that is necessary to obtain reliable estimates of fourty--two connectivity measures; to provide a global evaluation of the reconstructed connectivity networks, the authors use the Receiver Operating Characteristic (ROC) curve and the corresponding Area Under the Curve (AUC), computed through a sliding window approach: they divided the data in overlapping windows and average the AUC estimated from them. More recently, \cite{lietal16} identify the length of the recorded data as a critical parameter that can compromise intra--subject repeatability of MEG findings, when connectivity is studied in the source space, by means of five different connectivity measures, grouped depending on whether they analyze the phase or the amplitude of the source time courses.\\

\noindent
In the present study we investigate the impact of the length of the recorded data on three widely--used connectivity metrics, namely Imaginary Part of Coherency (IC) \citep{Nolte_04}, generalized Partial Directed Coherence (gPDC) \citep{baetal07} and frequency--domain Granger Causality (fGC) \citep{geweke_82}. 
Our aim is to characterize the reliability of these connectivity metrics in the source space, i.e. when they are computed between the neural source time courses estimated from EEG time series. We posit that there are at least three distinct contributions to error in these connectivity estimates: one comes from the finiteness of the data length; one from the fact that the source time courses are, in turn, estimated from EEG time series through an inverse procedure; the last one comes from the presence of biological noise, that provides an additional source of error.  In order to investigate separately the contributions of such error terms, we start by assuming knowledge of the exact source time courses, and investigate the effect of reducing the number of samples in the time series. Then we compute their EEG signals, affected by different levels of biological noise, and the corresponding estimated source time courses, computed with exact low resolution brain electromagnetic tomography (eLORETA) \citep{pascual_2007,pascual_2011}. We use surrogate data to assess the significance of the estimated connectivity values.\\ 

\noindent

%% file: methods.tex
\section{Materials and Methods}

Before describing in detail the materials and the procedures used in this work, we introduce the general methodological framework that has been used to assess the impact of the length of the input data on connectivity analysis performed in the source space. To this end, we rely on thee metrics, namely IC, gPDC and fGC. These three metrics have been chosen as they are representative of different classes of connectivity measures commonly used in the literature  \citep{basa14, faetal12, ei06}. Indeed, the former is an antisymmetric, bivariate measure computed from the Fourier Transform of the time series in input, and is thus completely data-driven. Conversely, gPDC and fGC are multivariate, directed measures which assume a MultiVariate AutoRegressive (MVAR) model for the data;
gPDC only measures direct connections, while fGC detects all (direct and indirect) connections.\\
In order to explore the differences among the considered connectivity measures, we devise the following simulation scheme in which a connectivity pattern is designed as four time series following a stable MVAR model, with information flowing only from the first to the second and to the third signal. This particular simulation scheme has been chosen because it allows us to compute also the theoretical values of the connectivity measures under investigation. Moreover, the common situation \citep{basc15} of the presence of a single signal influencing two different unconnected sources is simulated, and the difference in the behaviour of bivariate (IC) and multivariate (fGC and gPDC) measures when facing this situation is investigated. \\
We interpret these simulated time series as the time courses of four dipolar sources and we use them to generate synthetic EEG data with different levels of superimposed biological noise. Then, we perform connectivity estimation from sub--samples of increasing length extracted either from the original source time courses or from the neural activity reconstructed using eLORETA. The results obtained from the true source time courses are used as a reference for the results obtained from the EEG data. More specifically, using the surrogate data test we can compute how many of the estimated connectivity values turn out to be significant, i.e., large enough for rejecting the null hypothesis of no connectivity. We do the same for pairs of interacting sources (exploiting knowledge of the true connectivity values to asses if the values that pass the statistical test are actually relevant) and for pairs of independent sources (so as to quantify the ratio of false positives). 
Surrogate data for statistical significance are generated by means of two different procedures: Phase Randomization (PR) \citep{theiler_1992} and generation through an Autoregressive model (AR) \citep{schsch00}. Both of these surrogates share the auto-spectral density function of the original signals, the main difference being that PR data are generated by randomly changing the phase of the signals in input, whereas AR surrogate data are generated by making use of MVAR models. Finally, we compute a correlation coefficient between the estimated and the true connectivity values, as function of the frequency, to characterize the ability to recognise the frequency at which interaction occurs.
In the following, we first define the mathematical instruments used throughout the paper, i.e. the connectivity metrics, the statistical tests and the inverse model. Then, we present how we generate and analyse the data and eventually we describe the criteria used to evaluate the obtained results.\\
Everywhere in the paper, except where explicitly specified, the connectivity measures and the source reconstructions were computed using the Fieldtrip Matlab toolbox \citep{oostenveld_10}.

\subsection{Connectivity Metrics}\label{sec:connectivity}

Let $\{ \bm{x}(t_p) \}_{p=1}^P$ be a set of $N$ time series sampled at $P$ time points representing the activity of $N$ brain sources: 
\begin{equation}\label{eq:original_signals}
\bm{x}(t_p) = \left( x_1(t_p), \dots, x_N(t_p) \right)^T \quad p = 1, \dots, P.
\end{equation}

\noindent
In this study we consider three connectivity measures, all defined in the frequency--domain: Imaginary part of Coherency (IC), generalized Partial Directed Coherence (gPDC), and frequency--domain Granger Causality (fGC). We notice that these measures are strictly related to the ones used by \cite{haufe_13} for the simulated data available from the workshop titled \textit{Controversies in EEG source imaging}, held in August 2014 at the University of Electronic Science and Technology in Chengdu, China, with the aim of discussing the major issues in estimating brain activity and interaction properties from electrical potentials. All the simulations and data are available from the website http://neuroinformation.incf.org.\\

\paragraph{Imaginary Part of Coherency.} 
Imaginary Part of Coherency is a bivariate, antisymmetric measure of the linear relationship between two time series as function of frequency.\\
For each pair of signals $\left( x_i(t_p), x_j(t_p) \right), i,j = 1, \dots, N$, let $\left(\hat{x}_i(f_q),\hat{x}_j(f_q) \right)$ be the corresponding Fourier Transform. The Coherency \citep{nuetal97} between them is defined as the ratio between their cross--spectral density function and the product of their individual auto--spectral density functions, i.e. 
\begin{equation}
COH_{ij}(f_q) = \frac{S_{ij}(f_q)}{\sqrt{S_{ii}(f_q)S_{jj}(f_q)}}
\end{equation} 
where
\begin{equation}
S_{ij}(f_q) = <\hat{x}_i(f_q)\hat{x}_j(f_q)^*> ,
\label{eq:cdf}
\end{equation}
$x_j(f_q)^*$ being the complex conjugate of $x_j(f_q)$. Coherency is a complex valued quantity, the magnitude of which is strongly influenced by volume conduction and source leakage effects in MEG and EEG (M/EEG) which can induce spurious connectivity (\cite{Schoeffelen_2009}). However, \cite{Nolte_04} showed that for independent signals, only the real part of the cross--spectral density function can be non-zero, while its Imaginary part vanishes (a part from random fluctuations around zero). For this reason, in M/EEG it is common to use the Imaginary part of Coherency as a connectivity metric robust to spurious connectivity.
It is defined as \citep{Nolte_04} 
\begin{equation}
IC_{ij}(f_q) = \frac{\operatorname{Im}(S_{ij}(f_q))}{\sqrt{S_{ii}(f_q)S_{jj}(f_q)}} ~,
\label{eq:coh}
\end{equation}
$\operatorname{Im}(S_{ij}(f_q))$ being the imaginary part of $S_{ij}(f_q)$. 
We notice that the detection of functional coupling by the Imaginary part of Coherency is in fact influenced by the concurrent presence of independent sources at the frequency of interest, due to the normalization by the auto-spectral density functions; however, the effect of independent sources is a reduction of the value of $IC_{ij}(fq)$.

\noindent
From the computational viewpoint, the most delicate step for calculating $IC_{ij}(f_q)$ is the estimation of the cross-spectral density function, $S_{ij}(f_q)$, and of the auto--spectral density functions, $S_{ii}(f_q)$ and $S_{jj}(f_q)$. In the simulations below we use the Welch's method with a Hann window \citep{we67}: the measured signals are subdivided into $N_{win}$ epochs of fixed length (1 sec) and with an overlap of $50\%$. For each epoch and for each pair of signals, $i, j = 1, \dots, N$, we compute the FFT of  the time series $\left\{ x_i(t_p), x_j(t_p) \right\}_{p=1}^P$; the estimated cross--spectral density function is then given by Equation (\ref{eq:cdf}), where $< \cdot >$ stands for the average over epochs.

\paragraph{Generalized Partial Directed Coherence.} Generalized Partial Directed Coherence is a multivariate directed measure, whose definition is based on the assumption that the time series under investigation follow a stable MVAR process. Indeed, let us assume that the signals $\{ \bm{x}(t_p) \}_{p=1}^P$ can be  modelled through the following MVAR process
\begin{equation}
\bm{x}(t_p) = \sum_{k=1}^K A(k) \bm{x}(t_{p-k}) + \bm{\varepsilon}(t) ~, \label{eq:MVAR_proc}
\end{equation}  
where $K$ is the order of the process, $A(1), \dots, A(K)$ are proper coefficient matrices of size $N \times N$ and $\bm{\varepsilon}(t_p) = \left( \varepsilon_1(t_p), \dots, \varepsilon_N(t_p) \right)^T$ is a $N$--dimensional innovation process, which is a zero--mean, uncorrelated white noise process whit covariance matrix $\Sigma$. 
For each pair of signals, $i, j = 1, \dots, N$, the generalized Partial Directed Coherence from $\{ x_i(t_p) \}_{p=1}^P$ to $\{ x_j(t_p) \}_{p=1}^P$ is defined as \citep{baetal07} 
\begin{equation}
gPDC_{ij}(f_q) = \frac{\displaystyle \frac{1}{\sigma_{jj}} |\overline{A}_{ji}(f_q)|}{\sqrt{\displaystyle \sum_{n=1}^N \frac{1}{\sigma_{nn}^2} \overline{A}_{ni}(f_q)\overline{A}^*_{ni}(f_q)}}\label{eq:gPDC}
\end{equation}
where, for each $n = 1, \dots, N$, $\sigma^2_{nn}$ is the $n$--th diagonal element of $\Sigma$ and
\begin{equation}
\overline{A}_{ji}(f_q) = \delta_{ji} - \hat{A}_{ji}(f_q) 
\label{eq:trasf_A}
\end{equation} 
$\hat{A}_{ji}(f_q)$ being the Fourier Transform of the sequence of coefficients $\{ A_{ji}(k) \}_{k=1}^K$ and $\delta_{ji}$ the Kronecker delta function.\\
We observe that $gPDC_{ij}(f_q)$ is non--zero only if $A_{ji}(k) \neq 0$ for some $k$, i.e. only if the past of $x_i(t_p)$ directly influences $x_j(t_p)$. For this reason $gPDC_{ij}(f_q)$ is often referred to as a frequency--domain formulation of Granger Causality (see next paragraph for more details).\\
From the computational viewpoint, the most critical step for estimating the $gPDC_{ji}(f_q)$ is the fitting of the MVAR model. In the simulations below we estimate the model order $K$ by means of the Bayesian Information Criterion ($BIC$), which is a consistent estimator \citep{b_Ltkepohl_07}; specifically, we make use of the  MVGC toolbox \citep{barnett_14}.
The coefficient matrices, $A(1), \dots, A(K)$, and the covariance matrix of the innovations process, $\Sigma$, are estimated through the Levinson, Wiggins, Robinson (LWR) algorithm, implemented in the BSMART toolbox \citep{haykin_08,cui_08}.\\

\paragraph{Frequency--domain Granger Causality.}
Like gPDC, frequency--domain Granger Causality \citep{geweke_82} is a directed connectivity measure, whose definition is based on the assumption that the the signals $\{ \bm{x}(t_p) \}_{p=1}^P$ are drawn from a MVAR process.\\ 
An operational definition is as follows: given a pair of time series $\{ x_i(t_p), x_j(t_p) \}_{p=1}^P$, $i, j = 1, \dots N$, $\{ x_i(t_p) \}_{p=1}^P$ is said to Granger--cause $\{ x_j(t_p) \}_{p=1}^P$ if the past of the former process helps to predict the future of the latter \citep{wiener_56}.
In this study we consider the following formalization of Granger Causality in the frequency domain, which makes use of MVAR processes \citep{granger_69, geweke_82}.\\
Let $H(f_q)$ be the \textit{transfer matrix} of the MVAR model, i.e.
\begin{equation}
H(f_q) = \overline{A}(f_q)^{-1}\label{eq:transf_H}
\end{equation}
where $\overline{A}(f_q)$ is the $N \times N$ matrix whose $(i,j)$--th element is $\overline{A}_{ij}(f_q)$, as defined in Equation (\ref{eq:trasf_A}); then, for each pair of signals  $\left(x_i(t_p), x_j(t_p) \right), i,j = 1, \dots, N$, the frequency--domain Granger causality is given by  
\begin{equation}
	fGC_{ij}(f_q) = \ln \left( \frac{|S_{jj}(f_q)|}{|S_{jj}(f_q) - H_{ji}(f_q)(\Sigma_{ii} - \Sigma_{ij} \Sigma_{jj}^{-1} \Sigma_{ji}) H_{ji}(f_q)^* | } \right).
	\label{eq:fgc}
\end{equation}
Roughly speaking, $fGC_{ij}(f_q)$ quantifies the portion of the total auto--spectral density function $S_{jj}(f_q)$ that comes from a causal influence of $\{ x_i(t_p) \}_{p=1}^P$ over $\{ x_j(t_p) \}_{p=1}^P$ \citep{sch06}.\\
From a computational point of view, we estimate $fGC_{ij}(f_q)$ by first fitting an MVAR model, as described in the previous paragraph, and then estimating the transfer matrix $H(f_q)$ through (\ref{eq:transf_H}) and the cross--spectral density function by means of the following equation \citep{papi02}
\begin{equation}
	S(f_q) = H(f_q) \Sigma H(f_q)^*,
	\label{eq:cross_spectral}
\end{equation}
$H(f_q)^*$ being the Hermitian transpose of $H(f_q)$.

\subsection{Statical Test}\label{sec:stat_test}
In principle, all the three connectivity metrics defined above vanish if the input time series are independent. In practice, though, their empirical values are never exactly zero, even for independent signals, mainly due to the finite length of the data.
Therefore, once the connectivity values have been estimated, their significance has to be assessed, by quantifying the probability that those values could be generated in absence of connectivity.
In other words, we need to test the estimated values against the null hypothesis of no--connection among the signals. However, the distribution under the null hypothesis is usually unknown and a threshold of statistical significance has to be set by studying the asymptotic distribution of the connectivity measures \citep{baetal13} or by means of Monte Carlo methods such as the \textit{surrogate data approach} \citep{theiler_1992, schsch00}.\\ 
In this study, we use the latter approach. The idea of surrogate data is that of producing a large number, $N_{surr}$, of synthetic datasets which share given properties with the original signals, but in which the connectivity pattern under investigation is destroyed. For each pair of surrogate time series, $i, j = 1, \dots, N$, and for each frequency, $f_q$, the connectivity measures are computed and the distribution under the null hypothesis is approximated as the empirical distribution of the obtained values. Given a significance level $\alpha \in [0, 1]$, the threshold of statistical significance is then set at the $100 \ (1 - \alpha)$ percentile of the empirical distribution, i.e. $\tau_{ij}(f_q)$ is the lowest value such that 
\begin{equation}
\#\{\ \widetilde{M}^{(s)}_{ij}(f_q)\ s.t.\ |\widetilde{M}^{(s)}_{ij}(f_q)| > \tau_{ij}(f_q), \ s = 1, \dots, N_{surr}\ \} \leq (1-\alpha)\ N_{surr} ~,
\end{equation}
where $\widetilde{M}^{(s)}_{ij}(f_q)$ is the value of connectivity measure under investigation estimated for the $s$--th surrogate dataset and, given a set $A$, $\#A$ is the number of elements in A. Then, if the empirical value of the connectivity measure falls below the threshold, the probability that it has been generated in absence of connectivity is considered too high and the null hypothesis cannot be rejected. 
\\
Notice that, since the connectivity measures we
use in this study are frequency dependent, the thresholds are frequency dependent too.\\
Clearly, the first step is to define a way to generate surrogate data. This topic has been extensively studied in the recent years, so that rigorous methods have been developed, specifically devised to test the significance of given connectivity measures \citep{prth94, schsch96, faes10, liumo16}. Here we use two methods that are rather general and are appropriate for all the three connectivity metrics considered in this study. Both methods produce independent time series which share the auto--spectral density function with the original signals. The first one, i.e. \textit{phase--randomized surrogate data}, is the simplest one as it does not require model assumptions, whereas the second one, i.e. \textit{autoregressive surrogate data}, makes use of MVAR models.\\

\paragraph{Phase--randomized surrogate data.}
Let $\hat{\bm{x}}(f_q) = \left(\hat{x}_1(f_q), \dots, \hat{x}_1(f_q) \right)^T$ be the Fourier Transform of the original time series $\{ \bm{x}(t_p) \}_{p=1}^P$. \\
For each signal, $i = 1, \dots, N$, a \textit{phase--randomized} (PR) surrogate data\footnote{In literature, this type of surrogate data is also known as \textit{Fourier transform surrogate data} \citep{faetal04}}, $\{ \chi_i^{(PR)}(t_p) \}_{p=1}^P$ is generated as follows \citep{faetal04, theiler_1992}.\\
First for each frequency, $f_q$, we define 
\begin{equation}
\hat{\chi_i}^{(PR)}(f_q) := |\hat{x}_i(f_q)|\ e^{-i \varphi_i(f_q)}
\end{equation}
where $|\hat{x}_i(f_q)|$ is the modulus of $\hat{x}_i(f_q)$ and $\varphi_i(f_q)$ is randomly drawn from a uniform distribution on $[-\pi, \pi]$ under the constraint
\begin{equation}\label{eq:con_phi}
\varphi_i(-f_q) = - \varphi_i(f_q).
\end{equation} 
Secondly $\{ \chi^{(PR)}_i(t_p) \}_{p=1}^P$ is computed as the inverse Fourier Transform of $\hat{\chi}^{(PR)}_i(f_q)$. The constraint in equation (\ref{eq:con_phi}) is required to obtain a real valued time series.\\
From a computational viewpoint, we generate PR surrogate data by means of the eMVAR toolbox \citep{faetal13}.

\paragraph{Autoregressive surrogate data.}
For each signal, $i = 1, \dots, N$, we assume that $\{x_i(t_p) \}_{p=1}^P$ can be modelled as a unidimensional, autoregressive process of order $K_i$, coefficients $A_i(1), \dots A_i(K_i)$ and with covariance of the innovation $\sigma_i^2$. \\
An \textit{autoregressive} (AR) surrogate data \citep{faetal04, schsch00}, $\{ \chi^{(AR)}_i(t_p) \}_{p=1}^P$, is a realization of such AR process, i.e. for each $p=1, \dots, P$,
\begin{equation}
\chi^{(AR)}_i(t_p) = \sum_{k=1}^{K_i} A_i(k) \chi^{(AR)}_i(t_{p-k}) + \varepsilon_i(t_p)
\end{equation}
where $\varepsilon_i(t_p) \sim \mathcal{N}(0, \sigma_i^2)$.\\
From a computational viewpoint, first we estimate $K_i$ by means of the BIC criterion implemented in the MVGC toolbox \citep{barnett_14}; secondly we use FieldTrip to estimate the other parameters of the AR model and to generate $\{\chi^{(AR)}_i(t_p)\}_{p=1}^P$.

\subsection{Source reconstruction}\label{sec:inv}

Let $\bm{y}(t_p)$ be the EEG time series, recorded by $N_s$ sensors; we assume that the data have been produced by neural sources through a linear model of the form
\begin{equation}\label{forward}
\bm{y}(t_p) = G \bm{z}(t_p) + \bm{\eta}(t_p)
\end{equation}
where $G$ is the leadfield matrix, of size $N_s \times 3\/N_v$, $N_v$ being the number of voxels; $\bm{z}(t_p)$ is a vector of length $3 \/ N_v$ containing the three orthogonal components of the neural current at each voxel and $\bm{\eta}(t_p)$ is the measurement noise.\\
To reconstruct the source time courses $\bm{z}(t_p)$ from the recorded potential $\bm{y}(t_p)$ we use the exact low resolution brain electromagnetic tomography (eLORETA) \citep{pascual_2007,pascual_2011}. eLORETA belongs to the class of minimum norm solvers; specifically, the estimated source time courses are given by
\begin{align}
\tilde{\bm{z}}(t_p) & = \arg\min_{\bm{z}(t_p)} \left\{||\bm{y}(t_p) -  G \bm{z}(t_p) ||^2 + \lambda\ \bm{z}(t_p)^T\ W \ \bm{z}(t_p) \right\} \\
			   & = W^{-1}G^T \left(G W^{-1} G^T + \lambda I \right)^{-1}\ \bm{y}(t_p) \label{eq:eLORETA_exp}
\end{align}
where $\lambda$ is a regularization parameter and $W$ is a \textit{weight matrix}, of size $3N_v\ \times\ 3N_v$, defined in order to obtain zero localization error in the case of point-test sources and ideal noise--less condition. \cite{pascual_2007} shows that this can be achieved defining $W$ as a block--diagonal matrix
\begin{equation}
W = \left( \begin{array}{cccc}
W_1 &  0 & \dots & 0\\
0 &  W_2 & \dots & 0\\ 
\vdots & \ddots & \ddots & \vdots \\
0 & 0 & \dots & W_{N_v} 
\end{array} \right)
\end{equation}
where each block $W_v$, $v = 1, \dots, N_v$, has size $3 \times 3$ and satisfies
\begin{equation}
W_v^2 = G(\bm{r}_v)^T \left( G W^{-1} G^T + \lambda I \right)^{-1} G(\bm{r}_v) 
\end{equation}
$G(\bm{r}_v)$ being the leadfield matrix of the voxel centred in $\bm{r}_v$.

\subsection{Data simulation and analysis pipeline}\label{sec:analysis_pipeline}

\paragraph{Original source time courses.} We generate $N_d = 100$  dataset of length P = 10000 time points. Each dataset is composed by $N=4$ signals, referred to as the \textit{original} source time courses,
\begin{equation}
\bm{x}(t_p) = \left(x_1(t_p), x_2(t_p), x_3(t_p), x_4(t_p) \right)^T ~,
\end{equation}
drawn from a stable MVAR process of order $K=5$. The coefficient matrices of the process are defined in such a way that information flows only from $x_1(t_p)$ to $x_2(t_p)$ and to $x_3(t_p)$. Thus for each dataset we have
\begin{small}
\begin{equation}\label{eq:MVARmodel}
\left( \begin{array}{c}  x_1(t_p) \\ x_2(t_p) \\ x_3(t_p) \\ x_4(t_p) \end{array} \right) = 
\sum_{k=1}^K \left( \begin{array}{cccc} 
A_{11}(k) & 0 & 0 & 0 \\
A_{12}(k) & A_{22}(k) & 0 & 0 \\
A_{13}(k) &  0 & A_{33}(k) & 0 \\
0 &  0 & 0 & A_{44}(k) 
\end{array} 
 \right) 
\left( \begin{array}{c}  x_1(t_{p-k}) \\ x_2(t_{p-k}) \\ x_3(t_{p-k}) \\ x_4(t_{p-k}) \end{array} \right) +
\left( \begin{array}{c}  \varepsilon_1(t_p) \\ \varepsilon_2(t_p) \\ \varepsilon_3(t_p) \\ \varepsilon_4(t_p) \end{array} \right) 
\end{equation}
\end{small}
where the non zero coefficients, $A_{ij}(k)$, are drawn from a zero--mean normal distribution with variance of $0.01$, and the innovation $\bm{\varepsilon}(t_p) = \left(  \varepsilon_1(t_p), \varepsilon_2(t_p), \varepsilon_3(t_p), \varepsilon_4(t_p)  \right)^T$ is drawn from a standard Normal distribution ($\Sigma = I$).\\		
The model in equation (\ref{eq:MVARmodel}) is similar to the model proposed by \cite{haufe_13}\footnote{In order to simulate $\{ \bm{x}(t_p)\}_{p=1}^P$ we modified the Matlab code available at \url{http://neuroinformation.incf.org\ } }, the only difference being the presence of two more signals: (i) the third signal $x_3(t_p)$, that, like $x_2(t_p)$, is influenced by $x_1(t_p)$: the aim here is to have two non--directly connected brain areas with a common influence; (ii) the fourth signal $x_4(t_p)$, which is independent from the other three: the aim here is to have a control signal for detecting false positives.\\ 

\noindent
As a first step, we investigate the impact of the data length on the estimates of the connectivity metrics computed from the original source time courses. 
While this is not of practical use, because the original time courses are seldom available in practice, it provides a sort of upper bound to the quality
of the estimated values.
To do this, we subdivide each dataset, $\{ \bm{x}(t_p) \}_{p=1}^P$, in subsamples of increasing length. We choose an increment of $500$ time points for a total of $20$ subsamples, so that the $l$-th subsample contains $P_l = 500\ l$ time points. 
In practical terms, assuming a sampling frequency of $250 Hz$, each dataset $\{\bm{x}(t_p)\}_{p=1}^P$ will consist of $40$ seconds of data, the shortest subsample will contain 2 seconds of data and the $l$-th subsample will contain $2\ l$ seconds of data.
\\
For each subsample, and for each pair of signals $i, j = 1, \dots, N$, we estimate the value of the three connectivity measures (IC, gPDC and fGC) and generate surrogate data as described in Section \ref{sec:stat_test}; more specifically, we generate 100 realizations for each type of surrogate data, and choose a significance level $\alpha = 0.01$. As described in the previous section, we use these surrogate data to calculate the threshold of statistical significance corresponding to the $\alpha$ level.\\
Finally, for each dataset and for each subsample we compare the estimated connectivity measures with the corresponding thresholds: for each frequency, only the values that pass the threshold are considered significant.\\ 

\begin{figure}
\includegraphics[width=0.8\textwidth]{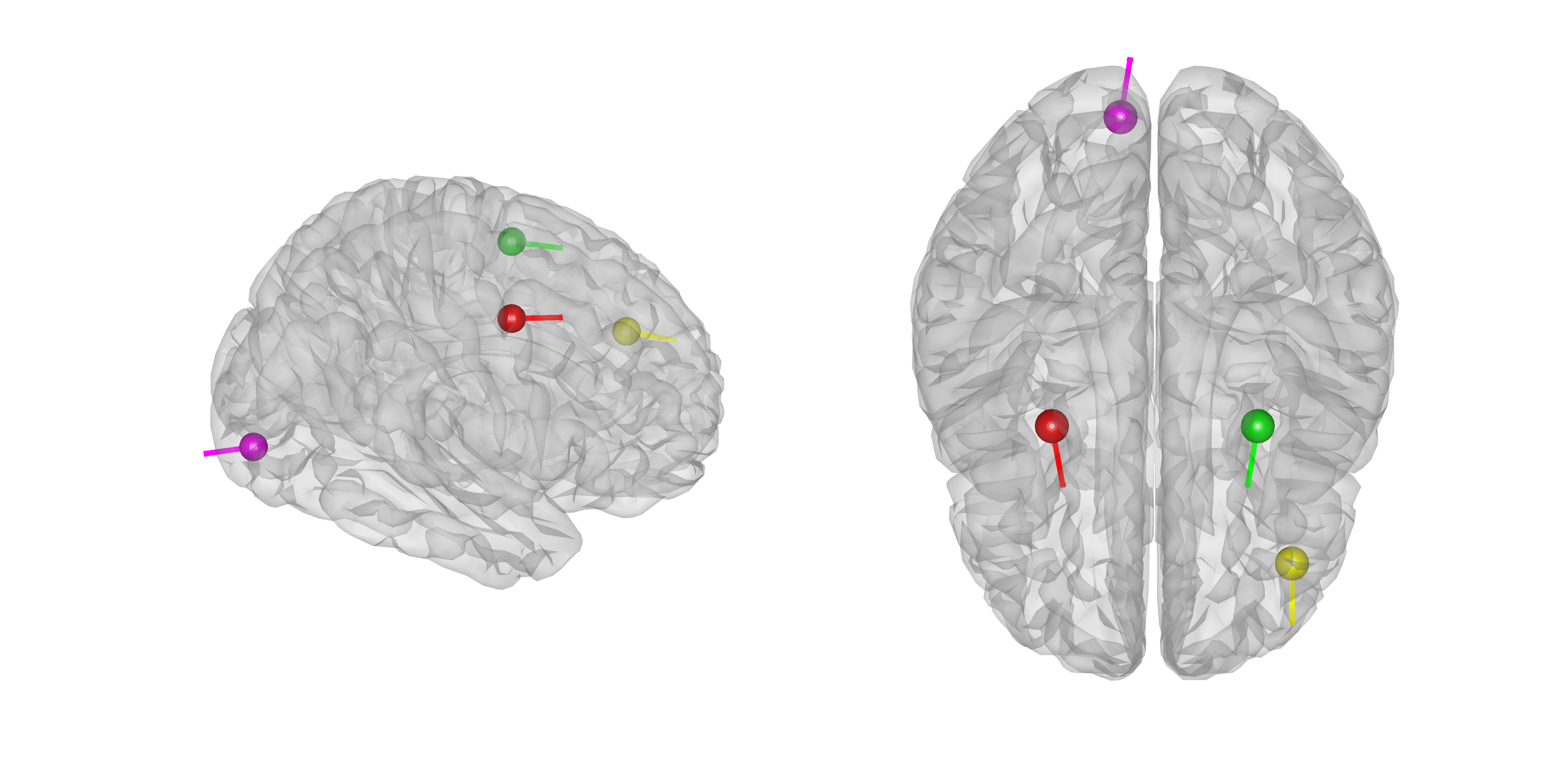}\caption{Dipole locations. Left: lateral view from the right. Right: sagittal view from the top, front at the bottom. Colored points and vectors represent the positions and orientations of the dipoles used to simulate the scalp potential $\{\bm{y}^{(i)}(t_p)\}_{p=1}^P$.}\label{fig:dip_pos}
\end{figure}

\paragraph{Estimated source time courses.}
In order to investigate the impact of the EEG forward and inverse model on the connectivity estimates, we simulate EEG datasets as follows. For each dataset of original source time courses $\{ \bm{x}(t_p) \}_{p=1}^P$, we interpret the four signals as the time courses of four dipolar sources 
that mimic the connectivity pattern observed in the stimulus driven control network \citep{cosh02}. In this network, a directed interaction between primary visual cortex and bilateral frontal areas has been reported. 
In our simulation dipole locations and orientations are set as in Figure \ref{fig:dip_pos}: the occipital purple source is located in the visual cortex and its time course is $\{ x_1(t_p) \}_{p=1}^P$; the green and red dipoles are placed in the proximity of the left and right frontal eye fields, being their time courses $\{x_2(t_p)\}_{p=1}^P$ and $\{x_3(t_p)\}_{p=1}^P$, respectively. An additional frontal source (yellow) models frontal activity uncoupled with the above system.\\

\begin{center}
\begin{figure}[ht]
\includegraphics[width=0.8\textwidth]{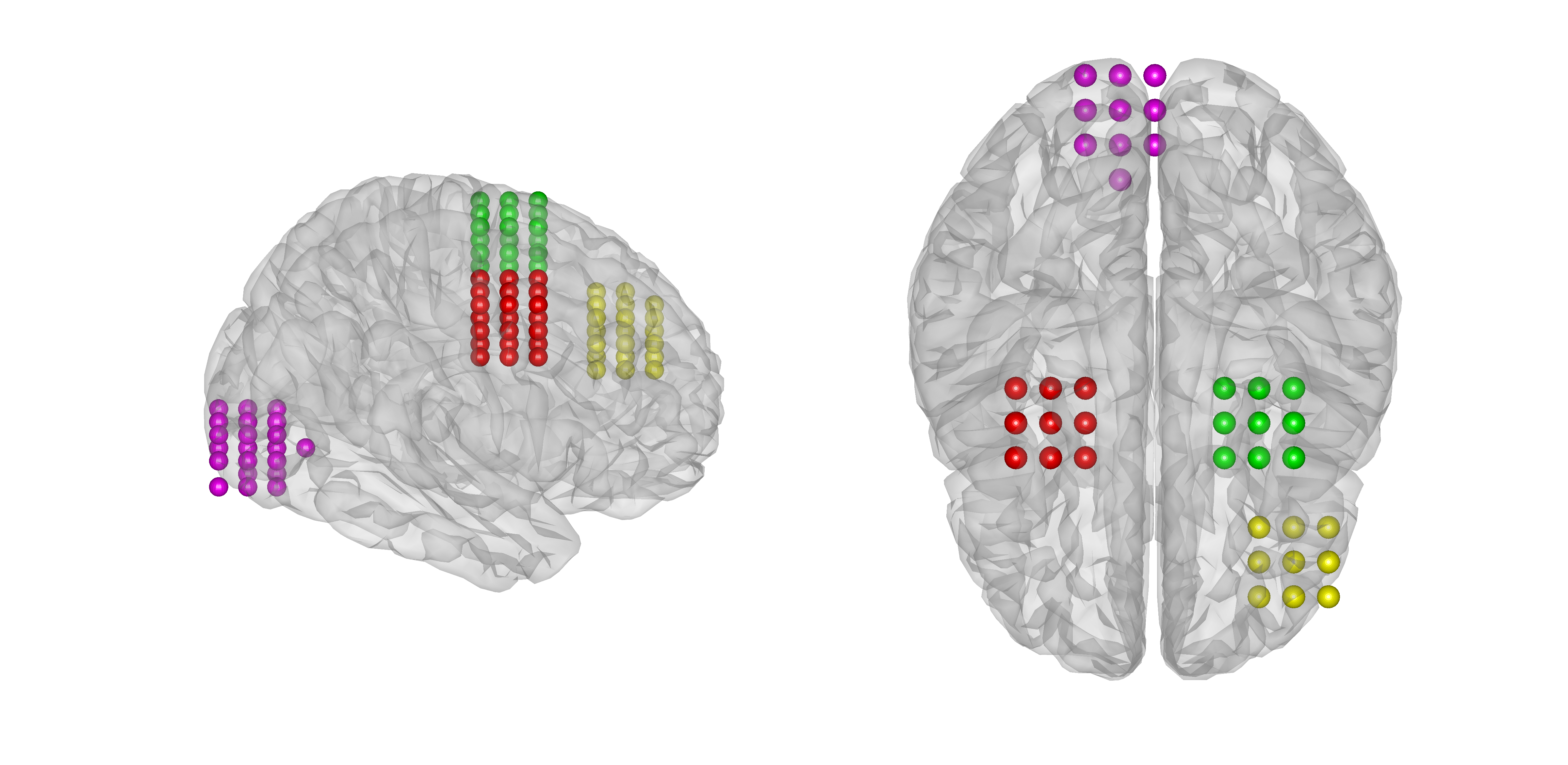}\caption{Region of interests. Left: lateral view from the right. Right: sagittal view from the top, front at the bottom. Each ROI is defined as the set of voxels that are within 2 cm from the location of the corresponding dipole of interest.}\label{fig:rois}
\end{figure} 
\end{center}

\noindent
We compute the scalp potential $\bm{y}^{(i)}(t_p)$ generated by these four sources according to \cite{noda05}: the forward model is based on three prolate spheroid shells that model a realistically shaped three--compartment (brain, skull and skin) head obtained from \cite{hoetal98}. A nose reference is used for the leadfield calculation \citep{haufe_13}. The presence of both \textit{biological} and \textit{sensor} noise is then simulated as follows.\\ 
Sensor noise $\bm{\eta}(t_p)$ is modelled by a standard normal distribution with fixed variance. Biological noise $\bm{y}^{(b)}(t_p)$ is simulated as the scalp potential generated by ten dipolar sources of random locations and orientations, and whose time courses follow each one a unidimensional autoregressive model\footnote{We observe that either the frontal yellow dipole or the 10 sources just described are not connected with anyone of the other dipoles of the networks. However, while the latter are used to simulate biological noise the former contributes to the scalp potential of interest and is used to evaluate the sensitivity of each connectivity measure to false positive.} 
 of order $10$. 
The three contributions are scaled in order to control the Signal to Noise Ratio (SNR): the final simulated EEG data is given by
\begin{equation}
\bm{y}(t_p) = (1-\gamma_1-\gamma_2) \frac{\bm{y}^{(i)}(t_p)}{||\bm{y}^{(i)}||} +
\gamma_1 \frac{\bm{y}^{(b)}(t_p)}{||\bm{y}^{(b)}||} + \gamma_2 \frac{\bm{\eta}(t_p)}{||\bm{\eta}||}
\end{equation}
where $\gamma_1$ and $\gamma_2$ weight the contribution of biological noise and sensor noise, respectively, and, given a time series $\{\bm{z}(t_p)\}_{p=1}^P$, $||\bm{z}||$ is the Frobenius norm of the matrix $\left(\bm{z}(t_1), \dots, \bm{z}(t_P) \right)^T$.
In the simulations below we set 
\begin{equation}
\gamma_2 = \frac{1}{8} \quad \quad \gamma_1 \in \left\{ 0,\ \frac{1}{4} (1-\gamma_2),\ \frac{1}{2} (1-\gamma_2) \right\}~~~,
\end{equation}
i.e. we consider three different noise levels.\\
From each scalp potential we reconstruct the brain activity $\{ \tilde{\bm{z}}(t_p) \}_{p=1}^P$ using eLORETA, Equation (\ref{eq:eLORETA_exp}). In particular, the inverse operator is computed with the same leadfield matrix used to generate the data, and setting the regularization parameter equal to the variance of the simulated sensor noise, i.e. 
\begin{equation}
\lambda = \left(\frac{\gamma_2}{||\bm{\eta}||}\right)^2 ~.
\end{equation}
Before studying connectivity we reduce the dimensionality of the problem by a two--step procedure. First we define four regions of interest (ROIs), each one containing the voxels that are within $2$ cm from one of the sources of interest (Figure \ref{fig:rois}); the activity of each ROI is estimated by summing up the activity of its voxels, thus reducing the overall dimension from $3N_v$ to $12$. Then we use Principal Component Analysis to project the activity of each area onto the direction of maximum power; we thus end up with four time series
\begin{equation}
\tilde{\bm{x}}(t_p) = \left(\tilde{x}_1(t_p), \tilde{x}_2(t_p), \tilde{x}_3(t_p), \tilde{x}_4(t_p) \right)^T, \quad p = 1, \dots, P
\end{equation} 
that estimate the original time courses $\{\bm{x}(t_p)\}_{p=1}^{P}$. \\

\noindent
To perform connectivity analysis from $\{ \tilde{\bm{x}}(t_p)  \}_{p=1}^P$, we proceed analogously to what we have done with the original time courses: we subdivide the reconstructed activity in sub--samples of increasing length, with the same lengths described in the previous paragraph. The values of the three connectivity measures, IC, gPDC and fGC, are estimated for each pair of reconstructed time courses, $\tilde{x}_i(t_p), \tilde{x}_j(t_p)$, $i, j = 1, \dots, N$, and their statistical significance is tested making use of PR and AR surrogate data.

\subsection{Evaluation of the performances}
Since the original source time courses are drawn from known MVAR processes, we can compute the \textit{theoretical values} of the connectivity measures under investigation. Indeed, first the transfer matrix $H(f_q)$ and the cross--spectral density function $S(f_q)$ can be computed from the model parameters through Equations (\ref{eq:transf_H}) and (\ref{eq:cross_spectral}). Then IC, gPDC and fGC are given by Equation (\ref{eq:coh}), (\ref{eq:gPDC}) and (\ref{eq:fgc}), respectively.\\
In the following simulations, these theoretical values are used to evaluate and compare the results obtained performing connectivity analysis from the subsamples of different length extracted either from the original datasets, $\{ \bm{x}(t_p) \}_{p=1}^P$, or from the reconstructed ROIs activity, $\{ \tilde{\bm{x}} (t_p) \}_{p=1}^P$.\\
First, we count how many estimated connectivity values pass the threshold of statistical significance. We do this for both the connected source pairs and for the
independent source pairs. For connected source pairs we provide histograms, where the counts are grouped based on the theoretical value of the connectivity measure. For independent source pairs we provide the ratio of false positives.\\
Then, we quantify the impact of the data length and (only for connectivity values computed from the estimated time courses) of biological noise by means of the correlation coefficient between the theoretical and the estimated values, as functions of frequency. More precisely we defined the correlation distance
\begin{equation}
\delta_{ij} := 1 - \frac{\sum_{q=1}^Q \left(M_{ij}(f_q) - \overline{M}_{ij} \right)\left(\tilde{M}_{ij}(f_q) - \overline{\tilde{M}}_{ij} \right)}{\sum_{q=1}^Q \left(M_{ij}(f_q) - \overline{M}_{ij} \right)^2 \sum_{q=1}^Q\left(\tilde{M}_{ij}(f_q) - \overline{\tilde{M}}_{ij} \right)^2}
\label{eq:correlation_distance}
\end{equation}
where $M_{ij}(f_q)$ and $\tilde{M}_{ij}(f_q)$ are the theoretical and estimated values of one of the connectivity measures between the pair of signals $\left(x_i(t_p), x_j(t_p) \right)$, $\overline{M}_{ij} = \sum_{q=1}^Q M_{ij}(f_q)$ and $\overline{\tilde{M}}_{ij} = \sum_{q=1}^Q \tilde{M}_{ij}(f_q)$.
We observe that $ 0 \leq \delta_{ij} \leq 2$ and $\delta_{ij} = 0$ when $\tilde{M}_{ij}(f_q) = \alpha M_{ij}(f_q)$ ($\alpha > 0$), i.e. when the empirical values correctly reproduce the behaviour over frequencies of the theoretical values.

%% file: results.tex
\section{Results}

\subsection{Threshold as a function of the data length}

\begin{figure}[ht!]
\includegraphics[width=\textwidth]{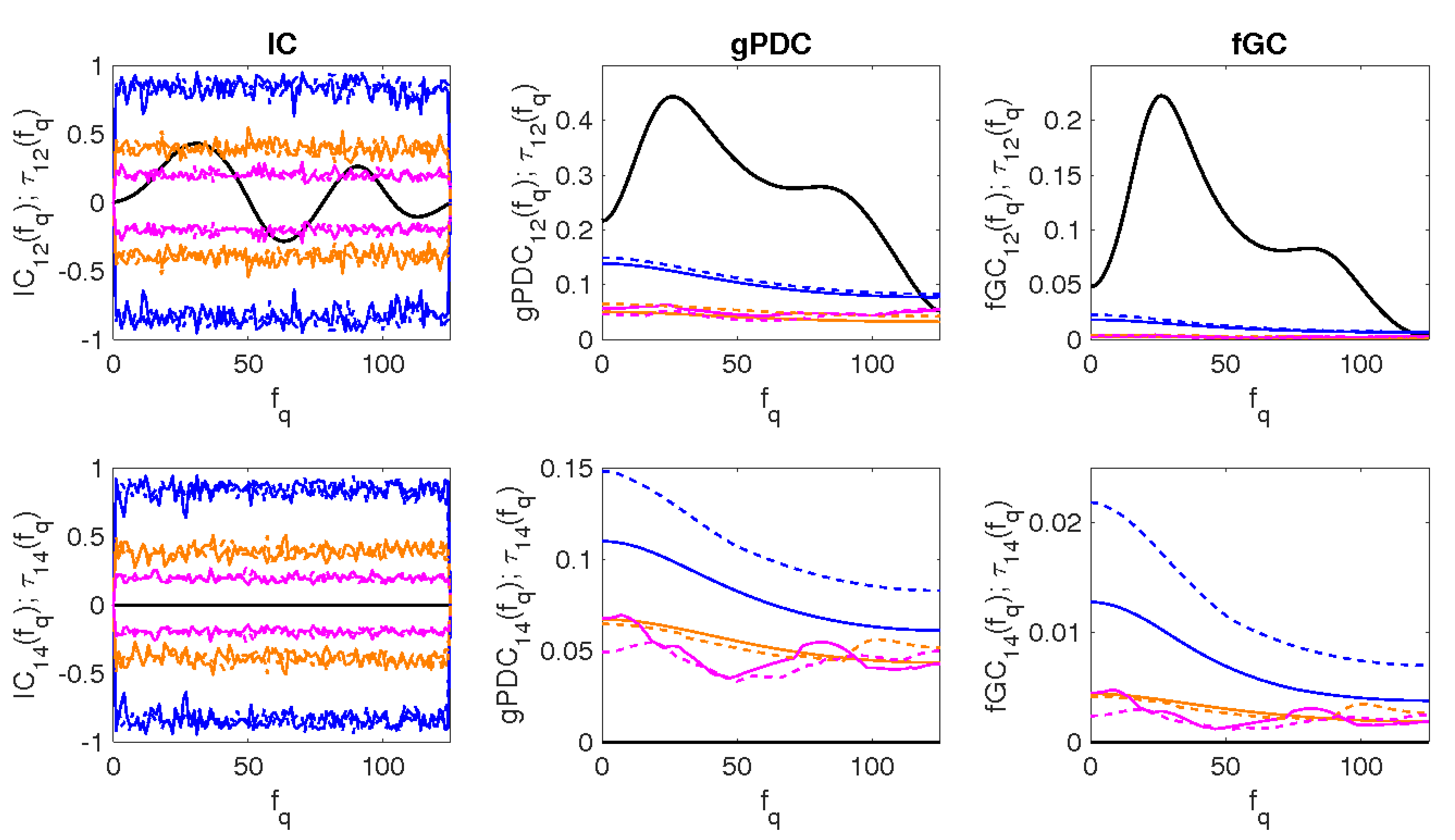}
\caption{Theoretical values and thresholds for a single dataset. For each connectivity measure we plot the theoretical value as a function of the frequency (black line) from $x_1(t_p)$ to $x_2(t_p)$ (first row) and from $x_1(t_p)$ to $x_4(t_p)$ (second row, independent signals). Then we superimpose the value of the threshold obtained from the sub--samples of length 500 (blue), 2500 (orange) and 10000 (magenta) time points. Dashed and solid lines are the thresholds obtained with PR and AR surrogate data, respectively. Notice the different scales on the y axes.}\label{fig:thresh_vs_true}
\end{figure}

\begin{figure}[H]
\includegraphics[width=1\textwidth]{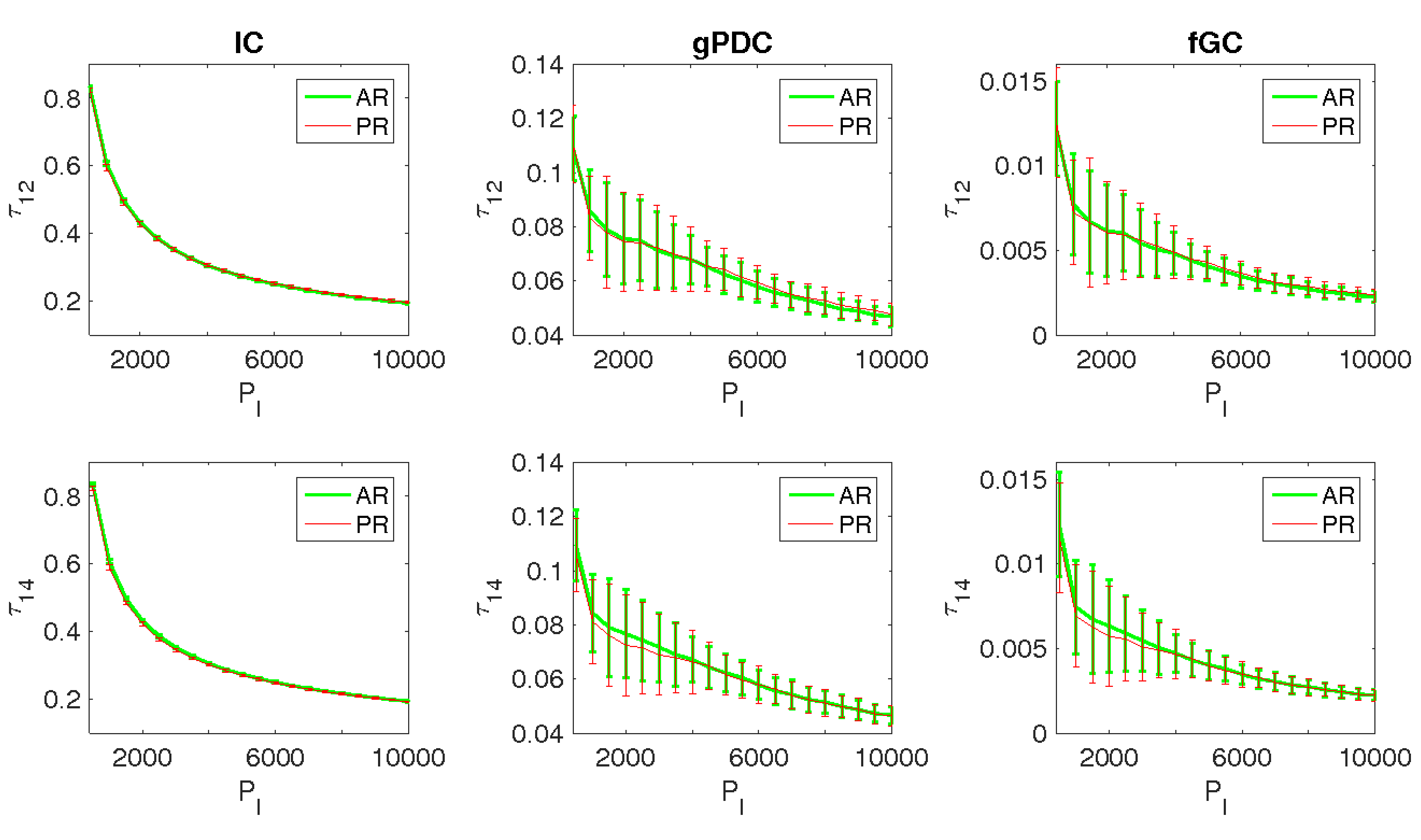}\caption{The threshold as a function of the length of the data: mean and standard deviation over frequencies and over test dataset for the pair $\left( x_1(t_p), x_2(t_p) \right)$ (first row) and for the pair $\left( x_1(t_p), x_4(t_p)\right)$ (second row). Thresholds have been obtained with PR (red line) and AR (green line) surrogate data.}\label{fig: thresh_vs_sslen}
\end{figure}

In Figure \ref{fig:thresh_vs_true}, we exemplify the behaviour of the threshold for statistical significance. The results here are obtained from a single realization of the original time courses, for illustrative purposes. For each connectivity measure we show two panels, one for the pair $(x_1(t_p),x_2(t_p))$ that has non--zero connectivity, one for the pair of independent signals $(x_1(t_p), x_4(t_p))$. The results from the other signal pairs are almost identical. In each panel we plot the theoretical values of the connectivity measure and the thresholds produced by PR and AR surrogate data as functions of the frequency. Different colors correspond to different sub--sample lengths. \\
As expected, the value of the threshold decreases as the number of samples in the time series increases, suggesting that lower connectivity values can be reliably estimated from longer time series. The two types of surrogate data produce almost identical thresholds for IC; in addition, the threshold values of IC appear to be frequency--independent. For gPDC and fGC the behaviour is quite different: the threshold do vary with the frequency, and the two types of surrogate data produce different thresholds, although there does not seem to be any systematic difference. Finally, the picture suggests that IC can hardly be estimated from the shortest sub--sample considered in this study, made of 500 time points, where the threshold is very close to 1; one can be more optimistic for gPDC and fGC, where the thresholds remain
lower than the true value even for the shortest sub--sample.\\
\noindent
To show that these behaviours do not depend on the specific choice of the source time courses, in Figure \ref{fig: thresh_vs_sslen} we plot mean and standard deviation of the threshold across the 100 datasets, as functions of the data length. This is in fact the result of a double averaging procedure: first, for each dataset we average across different frequencies; then, we average across the 100 datasets. As before, for each connectivity measure we show two panels, corresponding to the pair $(x_1(t_p),x_2(t_p))$ and $(x_1(t_p),x_4(t_p))$, respectively. 
The plots confirm that the thresholds decrease as the data length increases and that there is no systematic difference between PR and AR surrogate data. Moreover they show an almost zero variance for IC and a decreasing variance for gPDC and fGC. We notice that the higher variance of the thresholds for gPDC and fGC, compared to those for IC, is partly a consequence of the fact that they are frequency--dependent; in this sense, such higher variability is not necessarily a limitation, but may reflect a higher capability of capturing the main features of the original signals. In order to investigate these aspects, in the following section these thresholds are used to assess the statistical significance of the empirical values of each connectivity measure.

\subsection{Connectivity analysis from the original source time courses}\label{par:test_tadaset}
\begin{figure}[ht!]
\begin{center}
\subfigure[Imaginary part of Coherency]{\includegraphics[width=0.75\textwidth]{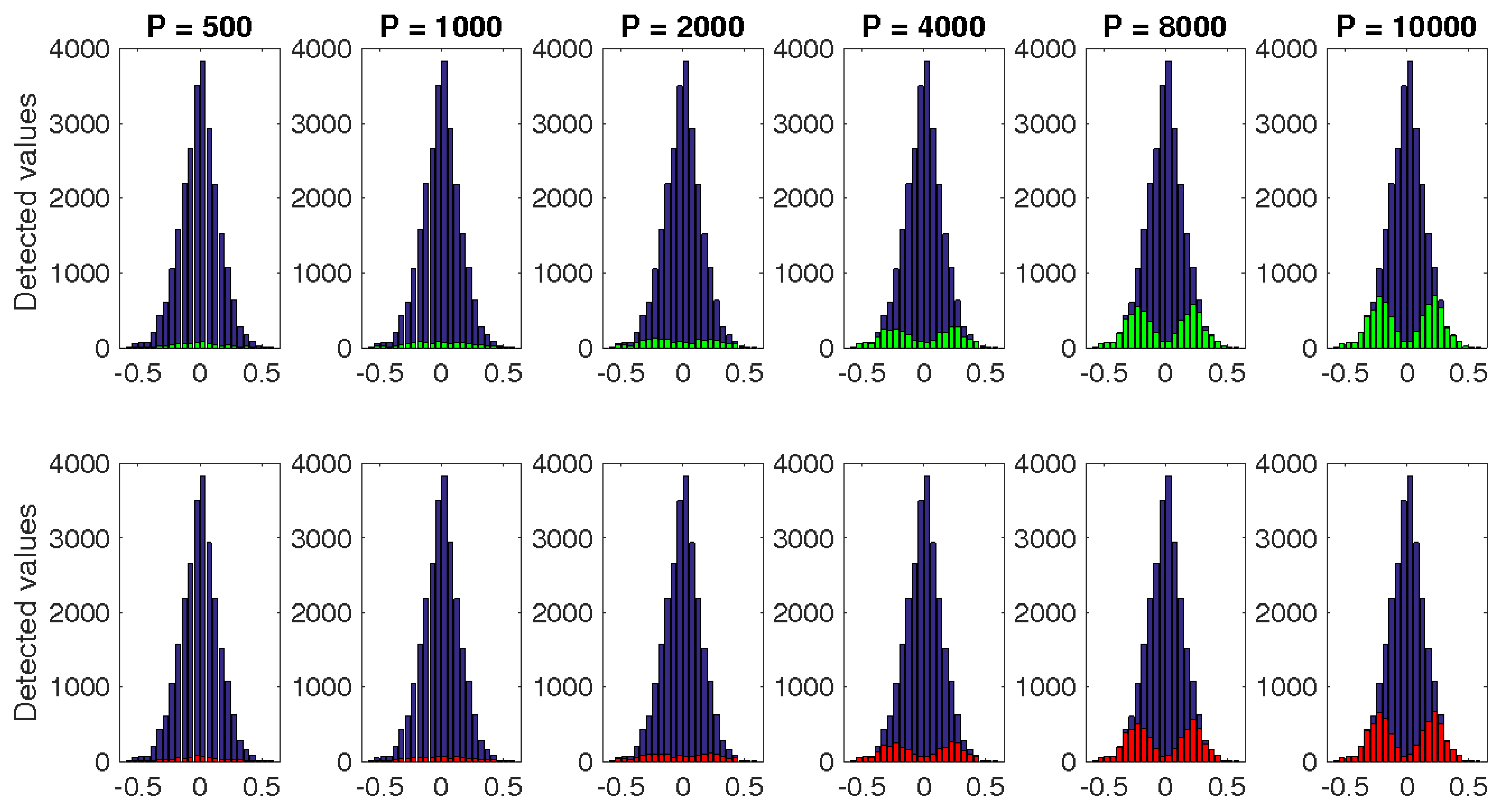}}
\subfigure[Generalized Partial Directed Coherence]{\includegraphics[width=0.75\textwidth]{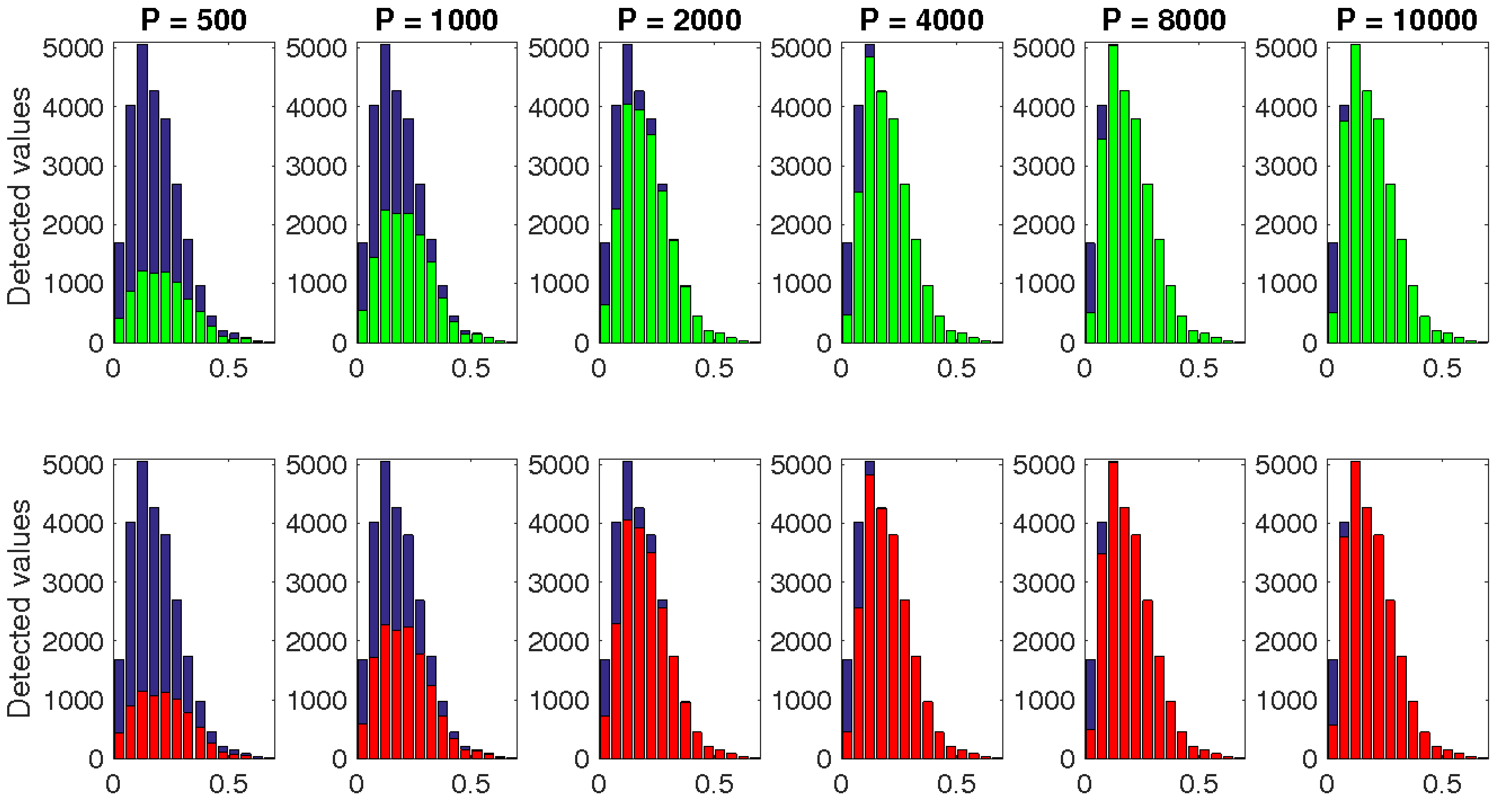}}
\subfigure[Frequency--domain Granger Causality]{\includegraphics[width=0.75\textwidth]{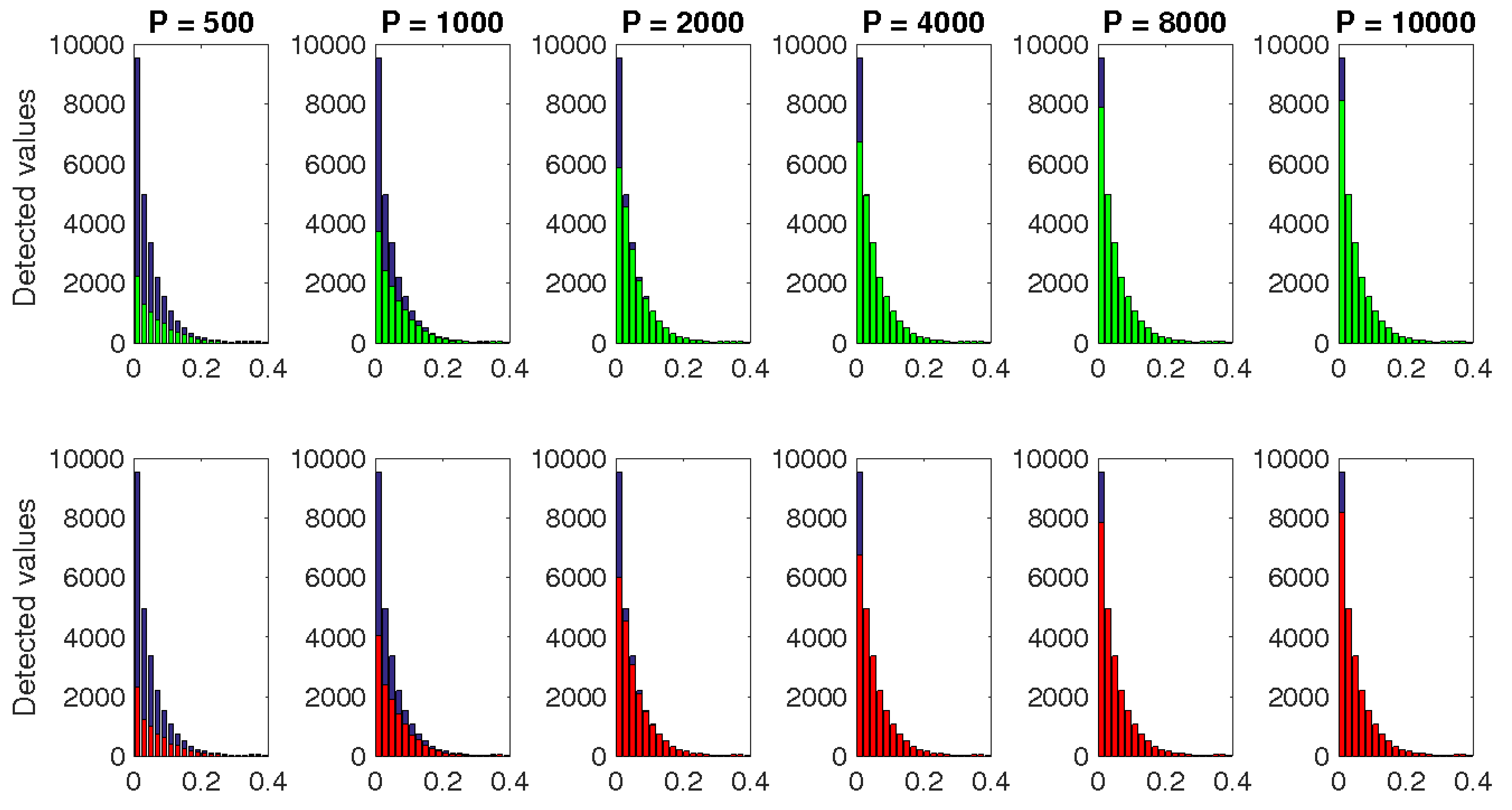}}
\end{center}\caption{Histrograms of the detected (red and green) and true (blue) values summed over different frequencies and datasets and over the two pairs $(x_1(t_p), x_2(t_p))$ and $(x_1(t_p), x_3(t_p))$. Different columns correspond to different sub--sample lengths, namely [500, 1000, 2000, 4000, 8000 10000] time points. Green and red bars represent the number of estimated values passing the statistical test with AR and PR surrogate data, respectively.}\label{fig:histogram}
\end{figure}

In this section we aim to investigate the capability of the different measures to detect the correct functional relationship among signals. As described in Section 2.4, for each sub--sample extracted from each dataset we estimate the empirical values of the connectivity measures and we assess their statistical significance by means of the threshold previously defined and analysed.\\
In Figure \ref{fig:histogram} we quantify the probability of correctly identifying the presence of connectivity as a function of the theoretical strength of such connectivity.
 
\begin{figure}[ht!]
\begin{center}
\includegraphics[width=\textwidth]{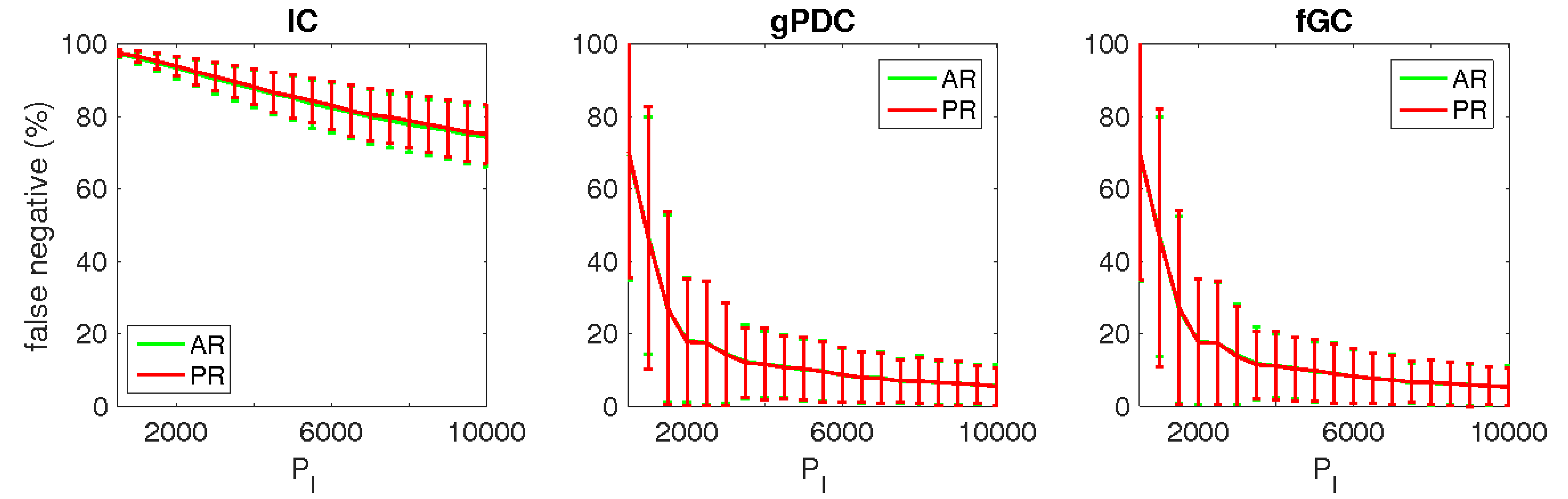}\\
\includegraphics[width=\textwidth]{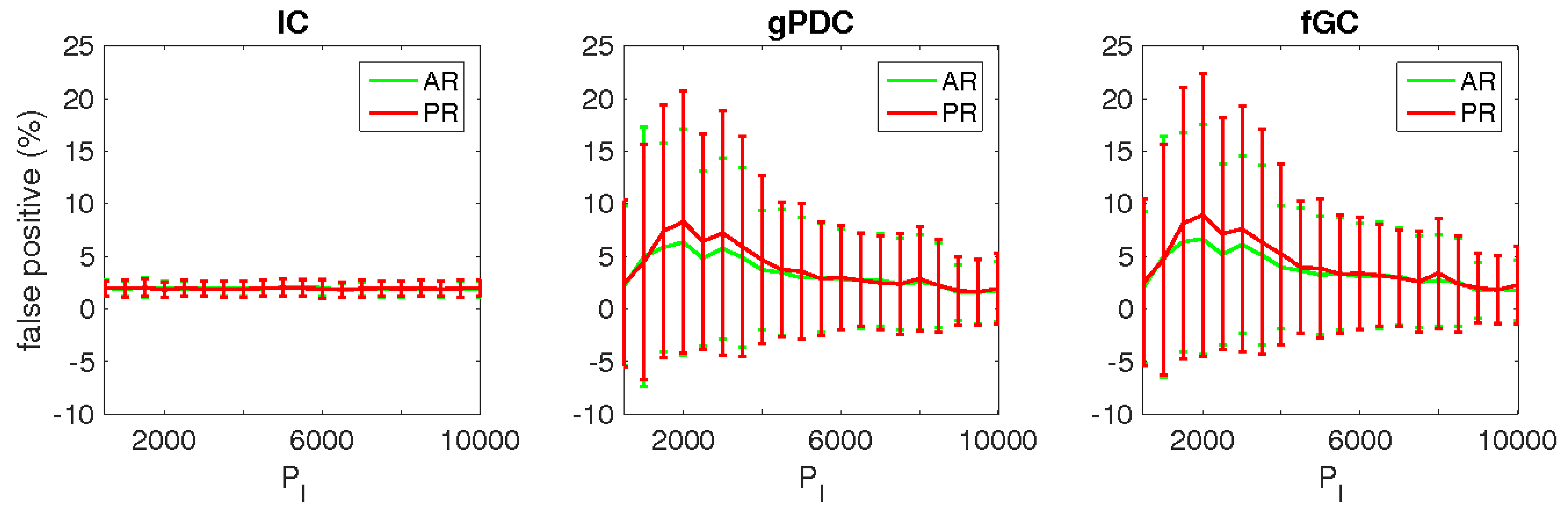}
\end{center}
\caption{Top/bottom: mean and standard deviation over datasets of the total percentage of false negatives/positives detected between the pairs of signals $(x_1(t_p), x_2(t_p))$ and $(x_1(t_p), x_3(t_p))$ / $(x_1(t_p), x_4(t_p))$, $(x_2(t_p), x_4(t_p))$ and $(x_3(t_p), x_4(t_p))$  as a function of the sub--sample length. The tresholds of statistical significance have been computed making use of AR (green line) and PR (red line) surrogate data. Each column correspond to a different connectivity measure. }\label{fig:fal_pas}
\end{figure}

\noindent
Indeed, we consider the (ordered) pairs of signals $(x_1(t_p), x_2(t_p))$ and $(x_1(t_p), x_3(t_p))$, which have non--null connectivity; for each measure, we compute the histogram of the theoretical connectivity values for all frequencies, summed up for all datasets and for the two pairs of signals; then, for each bar of such histogram we count how many of the corresponding empirical values pass the statistical test when the threshold is computed by using AR (green bar) or PR (red bar) surrogate data. In the figure we plot the results concerning the sub--samples of length [500, 1000, 2000, 4000, 8000, 10000] time points. The plots confirm that IC is the most ``conservative'' metric, i.e. IC values pass the threshold much more rarely than gPDC and fGC values. However, we also observe that, while above $P=2000$ the most of the high connectivity values are correctly recognised, below $P=2000$ the estimated values appear to pass the threshold more randomly, i.e. independently on the true underlying connectivity value. The results for gPDC and fGC are quite different: in fact, most estimated values pass the statistical test even at $P=2000$, and a large portion of them seems significant even at $P=500$.\\ 
In order to quantify the differences between the three measures, in Figure \ref{fig:fal_pas} we plot the percentage of false negatives as a function of the data length. This percentage is computed as follows. At each frequency a false negative is considered to occur if the empirical value of the connectivity measure does not pass the statistical test, even though the corresponding theoretical value is non--zero. Then we sum up the number of false negatives over frequencies and over the two pairs of signals $(x_1(t_p), x_2(t_p))$ and $(x_1(t_p), x_3(t_p))$. The Figure confirms our previous considerations. Indeed, for all the connectivity measures the number of false negatives decreases for increasing data length. Moreover, the false negative percentage for IC is systematically higher.\\
Clearly, these results feature a higher sensitivity of gPDC and fGC, i.e. a higher capability of the two connectivity measures to recognize the presence of connectivity, which in turn may also cause a higher number of false positives. To investigate this aspect, we evaluate the false positive ratio by making use of the pairs of unconnected signals $(x_1(t_p), x_4(t_p))$, $(x_2(t_p), x_4(t_p))$ and $(x_3(t_p), x_4(t_p))$. 
In the second row of Figure \ref{fig:fal_pas}, for each connectivity measure we plot the mean and the standard deviation over datasets of the percentage of false positives as a function of the sub--sample length. This percentage is computed by summing up the number of empirical values that pass the statistical test when the connectivity measures are computed between the three aforementioned pairs of signals. Please notice that, here and in figures below, the standard deviation has been plotted symmetrically with respect to the mean, though clearly the false positive ratio cannot assume negative values. The figure shows that for each connectivity measure the results obtained with either PR or AR surrogate data are almost identical. On the other hand, gPDC and fGC seem to be much more affected by false positives than IC. Indeed, while for IC the percentage of false positives remains very low for all the sub--sample lengths, for gPDC and fGC such percentage, and in particular the standard deviation across dataset, is considerably higher; while such percentage decreases as the sub--sample length increases, it never gets quite as low as that of IC.

\subsection{Common input problem}
\begin{figure}[ht!]
\begin{center}
\includegraphics[width=\textwidth]{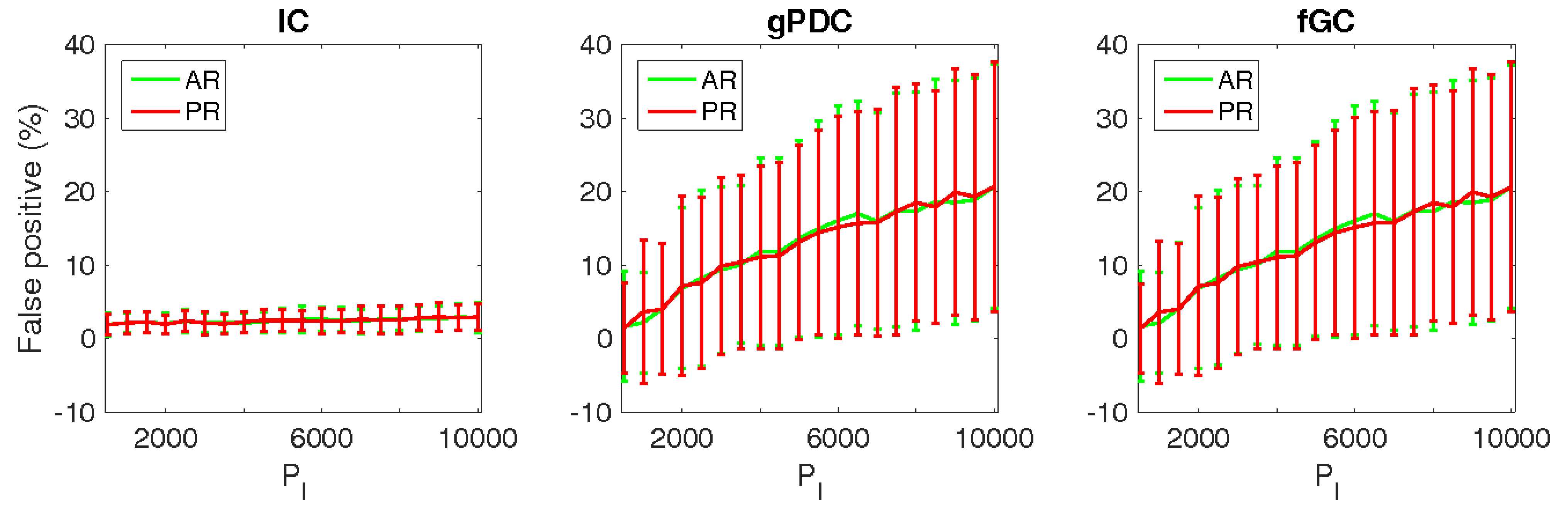}
\end{center}
\caption{Mean and standard deviation over the dataset of the total percentage of false positives between the order pair of signals $(x_2(t_p), x_3(t_p))$ and $(x_3(t_p), x_2(t_p))$ when the connectivity analysis is performed without considering the signals $x_1(t_p)$ and $x_4(t_p)$.}\label{fig:common_input}
\end{figure}

As described in Section \ref{sec:analysis_pipeline}, the original source time courses are simulated in such a way that there is no direct connection between $x_2(t_p)$ and $x_3(t_p)$; however, both are influenced by $x_1(t_p)$, the common input, and this induces some correlation between $x_2(t_p)$ and $x_3(t_p)$. The three connectivity metrics behave differently in this respect.
Since IC is a bivariate measure, there is no way to explicitly account for the presence of the common input in the estimation process. As a  consequence, the theoretical value of $IC_{23}(f_q)$ is non--zero; but, since the two sources are not actually connected, all the estimated values passing the threshold are to be regarded as false positives. In Figure \ref{fig:common_input}, left panel, we plot the false positive rate for the pair $(x_2(t_p), x_3(t_p))$: the values are somewhat larger with respect to those of purely independent signals (cfr. Figure \ref{fig:fal_pas}), but not dramatically so.
For gPDC and fGC, instead, there are two possibilities depending on whether the source $x_1(t_p)$ is taken into account or missed, due to, e.g., poor visibility in the data, as in the case of a radial source in MEG. If the common input is part of the MVAR model that is used in the estimation process, then the rate of false positives for the pair $(x_2(t_p),x_3(t_p))$ is the same of that for independent signals, depicted in Figure \ref{fig:fal_pas}. If, however, $x_1(t_p)$ is not part of the MVAR model, gPDC and fGC detect a spurious connectivity from $x_2(t_p)$ to $x_3(t_p)$ or vice versa, and the rate of false positives increases substantially. In Figure \ref{fig:common_input} we plot the rate of false positives obtained by neglecting the common input, i.e., by applying the analysis pipeline (MVAR model estimation, connectivity estimation, surrogate data generation) to the pair $(x_2(t_p),x_3(t_p))$.
Remarkably, the rate of false positives increases with the data length, because the spurious connectivity becomes more and more evident in the data.

\subsection{Connectivity analysis from the estimated source time courses}
\begin{figure}[H]
\begin{center}
\subfigure[Step 1: source reconstruction]{\includegraphics[width=0.8\textwidth]{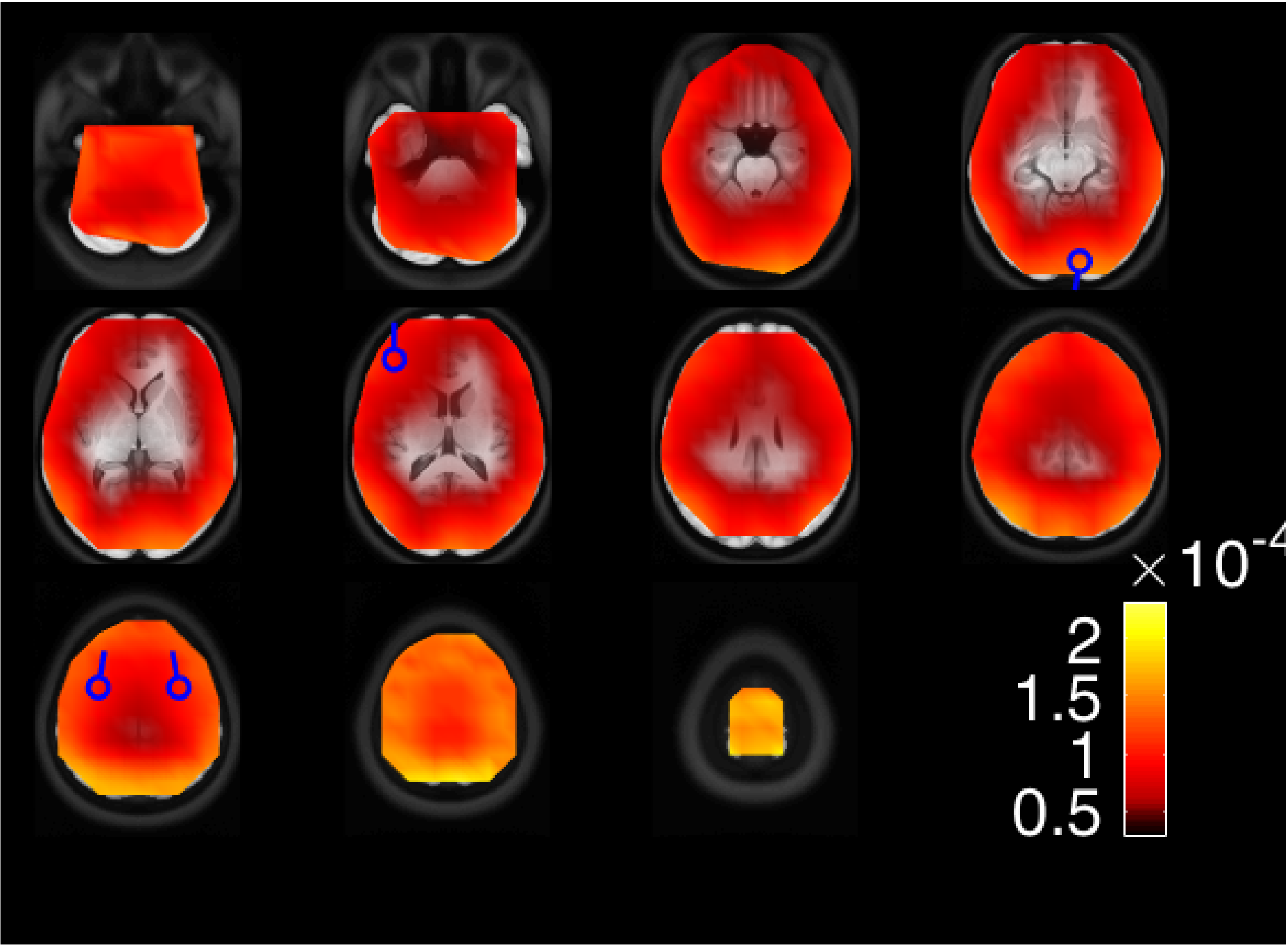}}\\
\subfigure[Step 2: dimensionality reduction]{\includegraphics[width=0.6\textwidth]{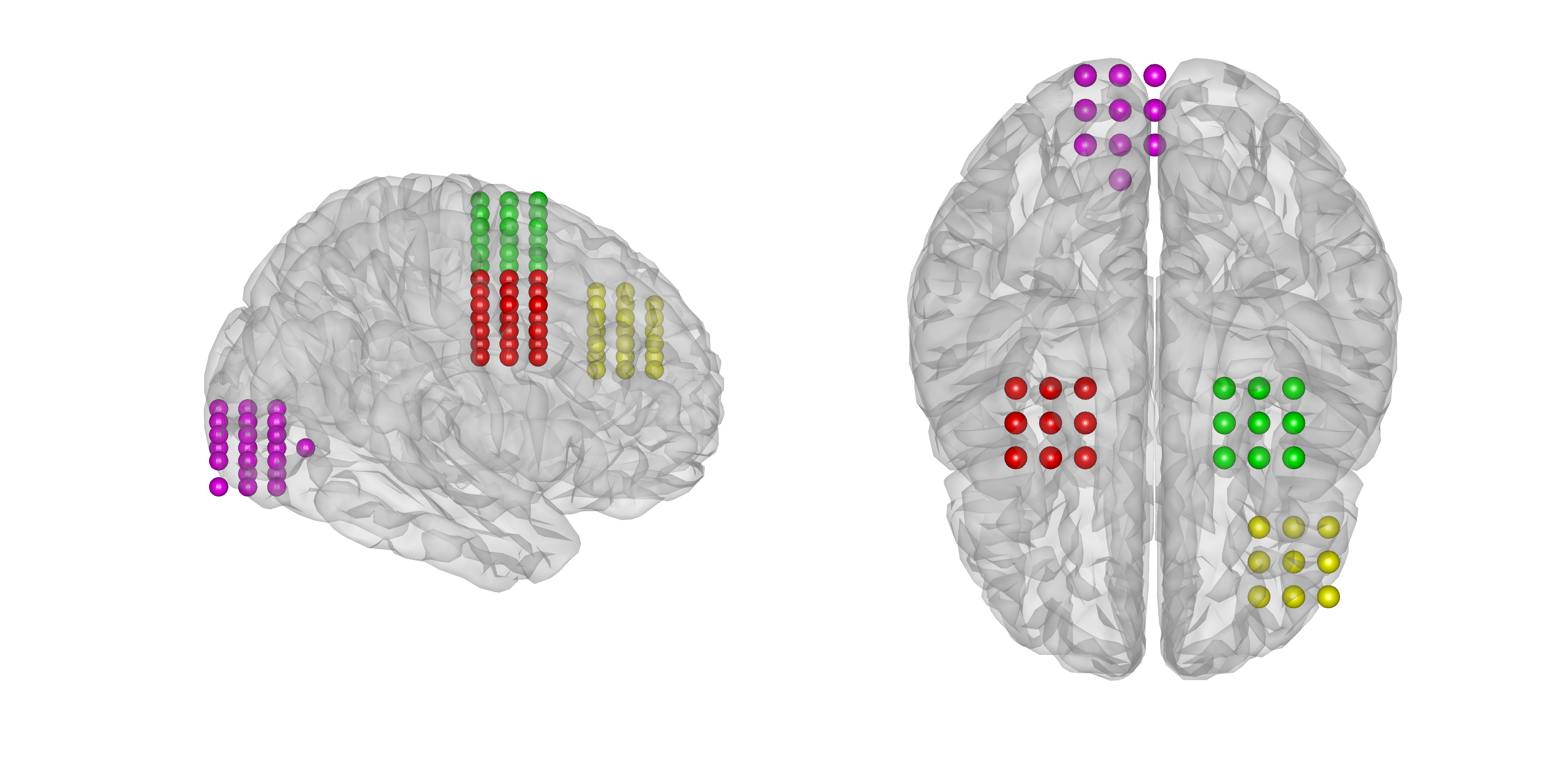}}\\
\subfigure[Step 3: estimation of the connectivity measures]{\includegraphics[width=\textwidth]{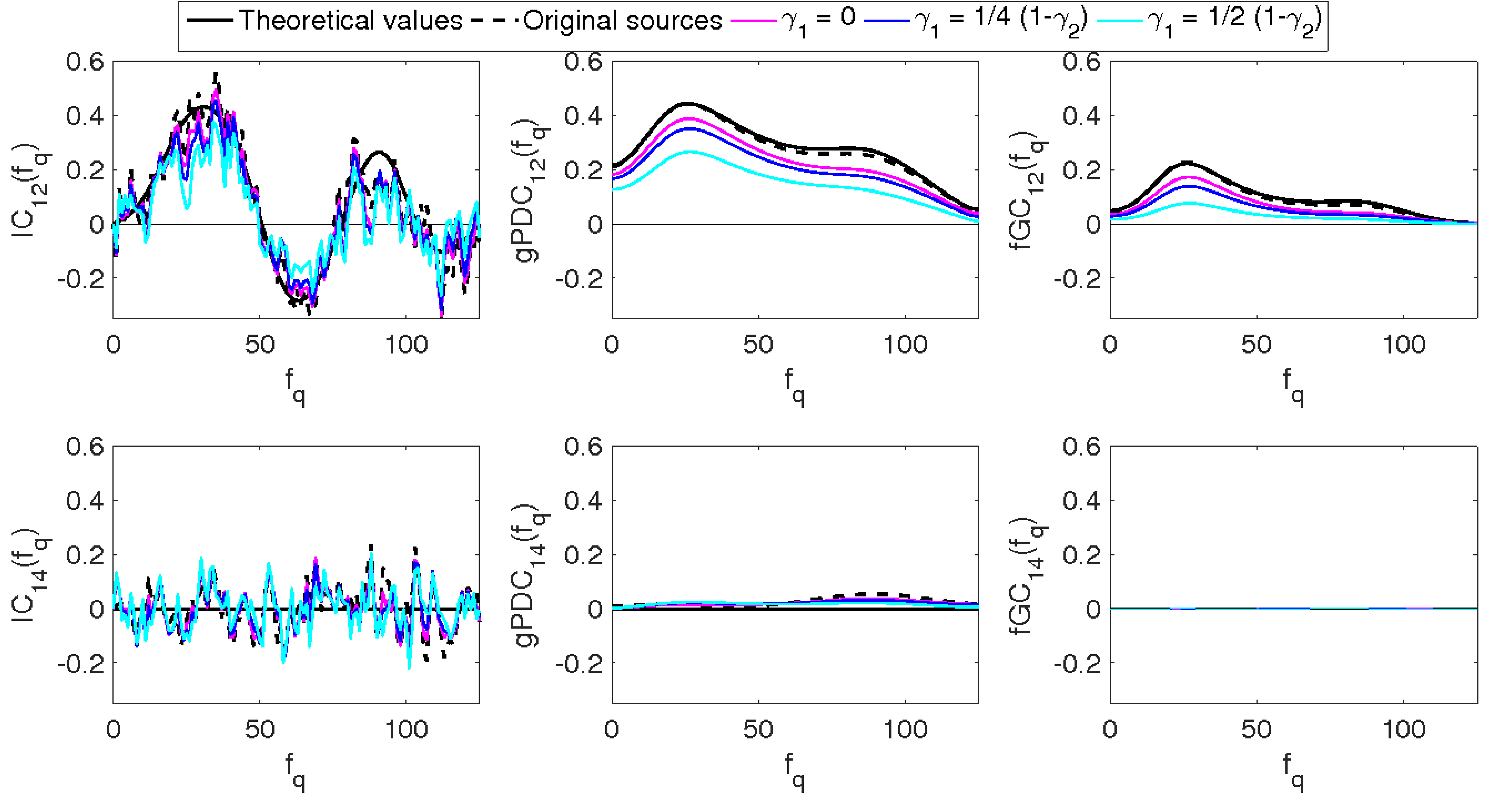}}
\end{center}\caption{Analysis pipeline for the simulated EEG data. Fixed one of the original source time courses $\{ \bm{x}(t_p)\}_{p=1}^P$, in the top panel, we plot the voxelwise amplitude averaged over time of the neural sources reconstructed from the EEG data generated setting $\gamma_1 = \frac{1}{4} \left(1-\gamma_2 \right)$, where $\gamma_2=\frac{1}{8}$. Blue circles and lines represent locations and orientations of the dipoles of interest. In the second panel, we plot the ROIs defined to reduce the number of time series used for the connectivity analysis. 
Eventually, in the last two rows we show the estimated connectivity measures for two pairs of ROIs. In each panel we plot the theoretical values of the measure (black line) and the values estimated from the original source time courses (dotted line) and from the reconstructed source activities obtained from EEG data simulated with different values of $\gamma_1$ (coloured lines). We observe that when there is no connectivity the computed values of fGC are close to zero in all the conditions (bottom right panel).}
\label{fig:source_amplitude}
\end{figure}

\begin{figure}[H]
\begin{center}
\subfigure{\includegraphics[width=0.8\textwidth]{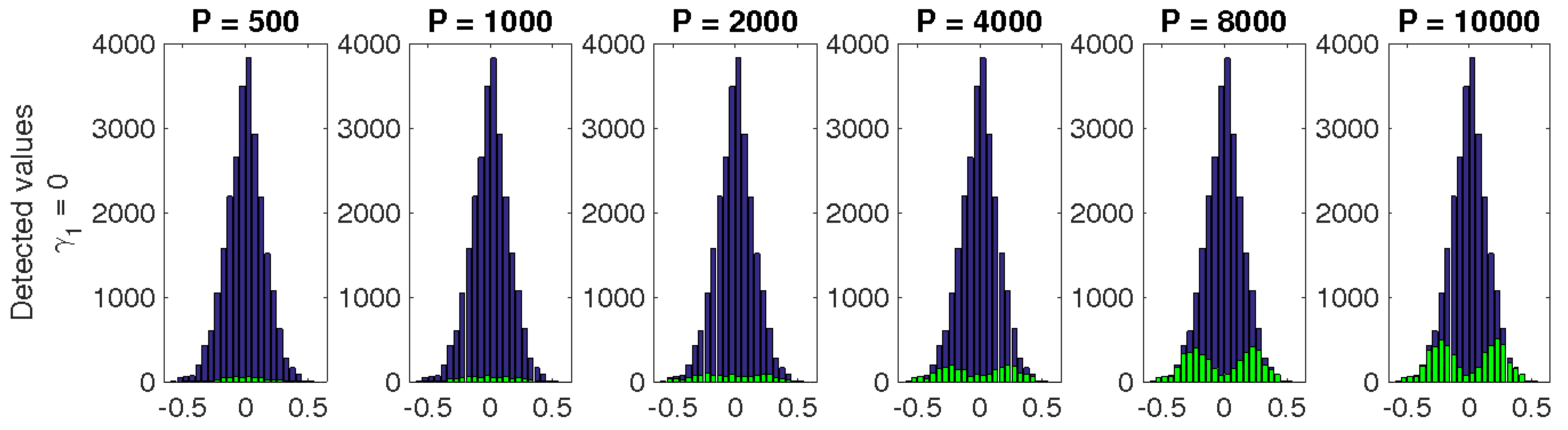}}
\subfigure{\includegraphics[width=0.8\textwidth]{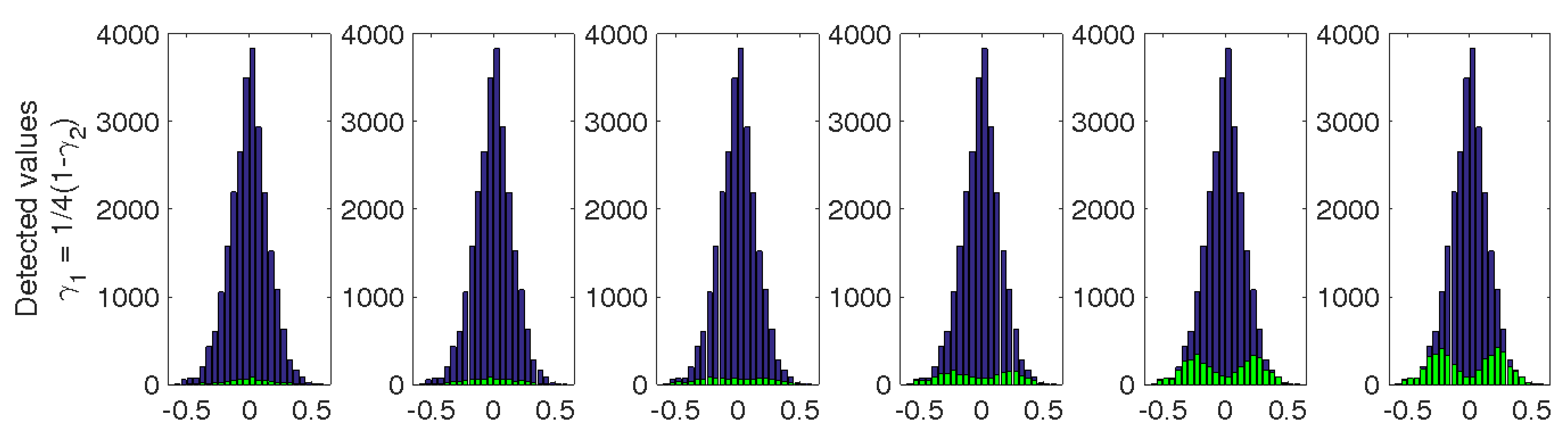}}
\subfigure[Imaginary part of Coherency]{\includegraphics[width=0.8\textwidth]{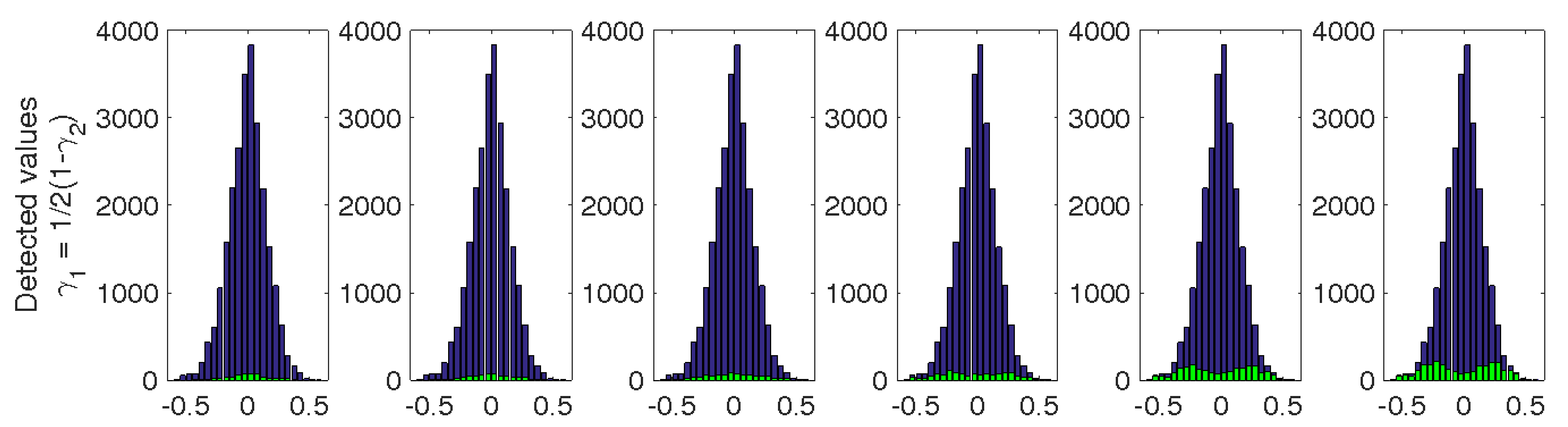}}
\end{center}
\caption{Histograms of the statistically significant values (green bar) grouped according to the theoretical values (blue bars), summed 
over different frequencies and datasets and over the two pairs of time series $(\tilde{x}_1(t_p), \tilde{x}_2(t_p))$ and $(\tilde{x}_1(t_p), \tilde{x}_3(t_p))$ reconstructed from EEG data with different levels of biological noise. Different columns correspond to different sub--sample lengths, namely [500, 1000, 2000, 4000, 8000 10000] time points. The statistical test is performed by means of AR surrogate data.}
\label{fig:hist_ic}
\end{figure}

\begin{figure}
\begin{center}
\subfigure{\includegraphics[width=0.8\textwidth]{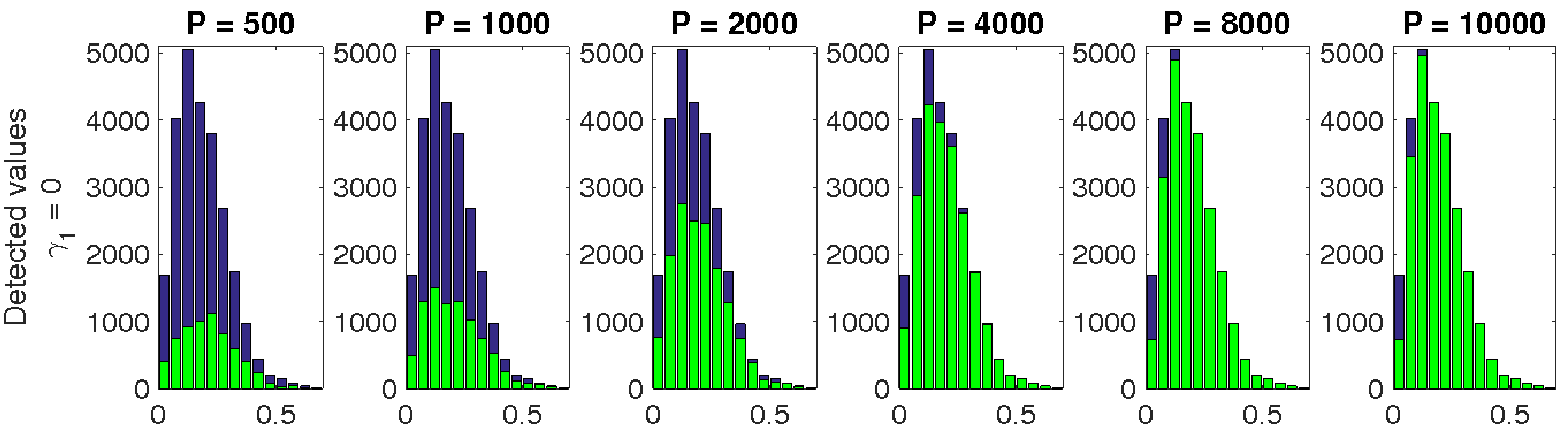}}
\subfigure{\includegraphics[width=0.8\textwidth]{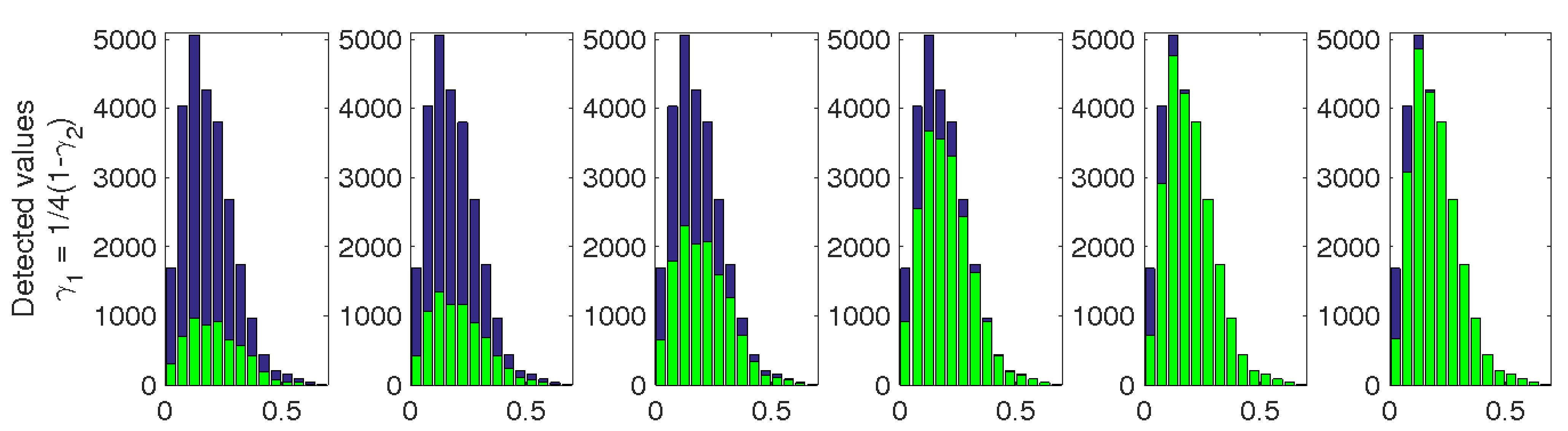}}
\subfigure[Generalized Partial Directed Coherence]{\includegraphics[width=0.8\textwidth]{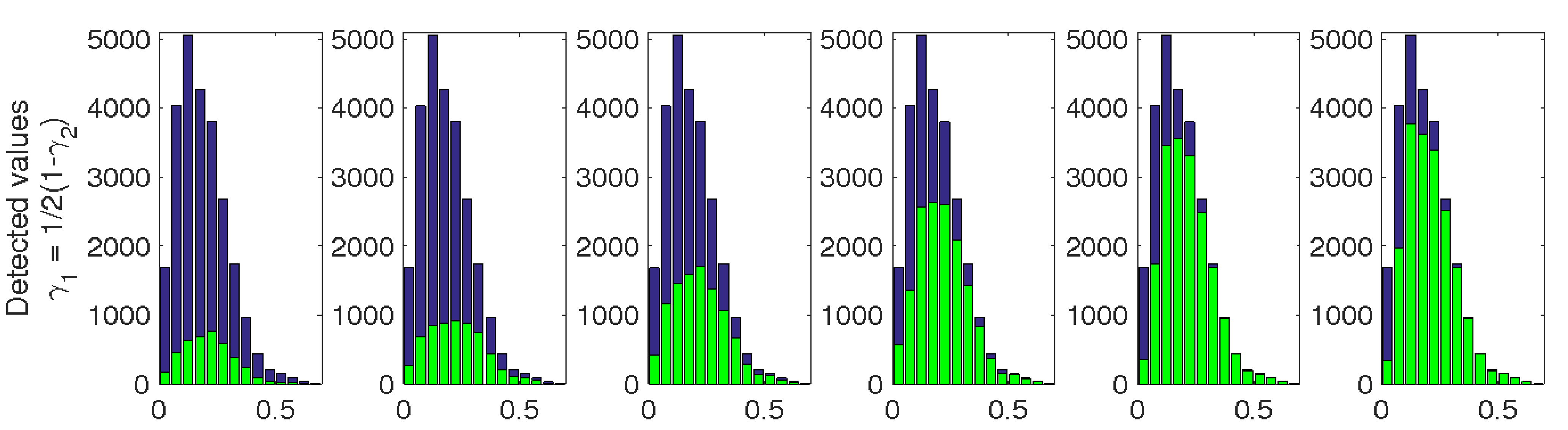}}
\end{center}
\caption{Histograms of the statistically significant values obtained as described in Figure \ref{fig:hist_ic}}
\label{fig:hist_pdc}
\end{figure}

\begin{figure}[H]
\begin{center}
\subfigure{\includegraphics[width=0.8\textwidth]{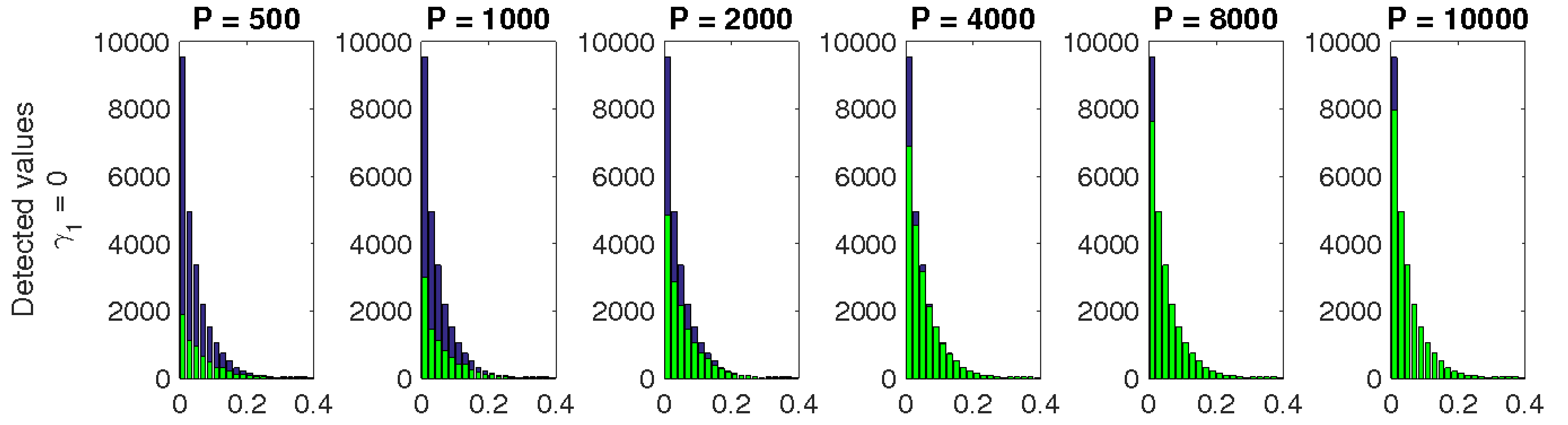}}
\subfigure{\includegraphics[width=0.8\textwidth]{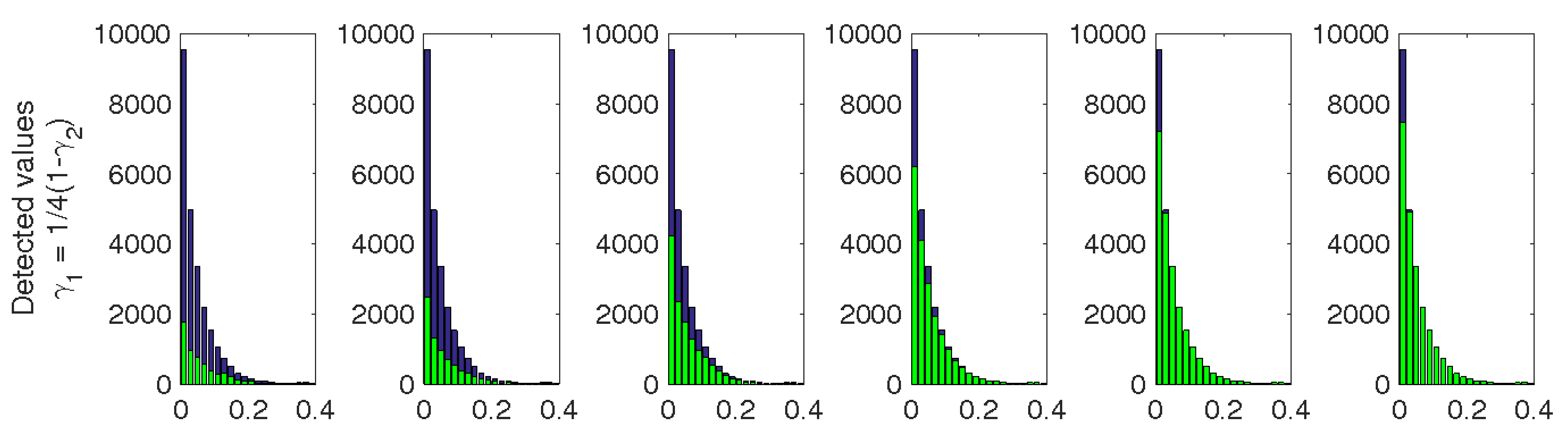}}
\subfigure[Frequency--domain Granger Causality]{\includegraphics[width=0.8\textwidth]{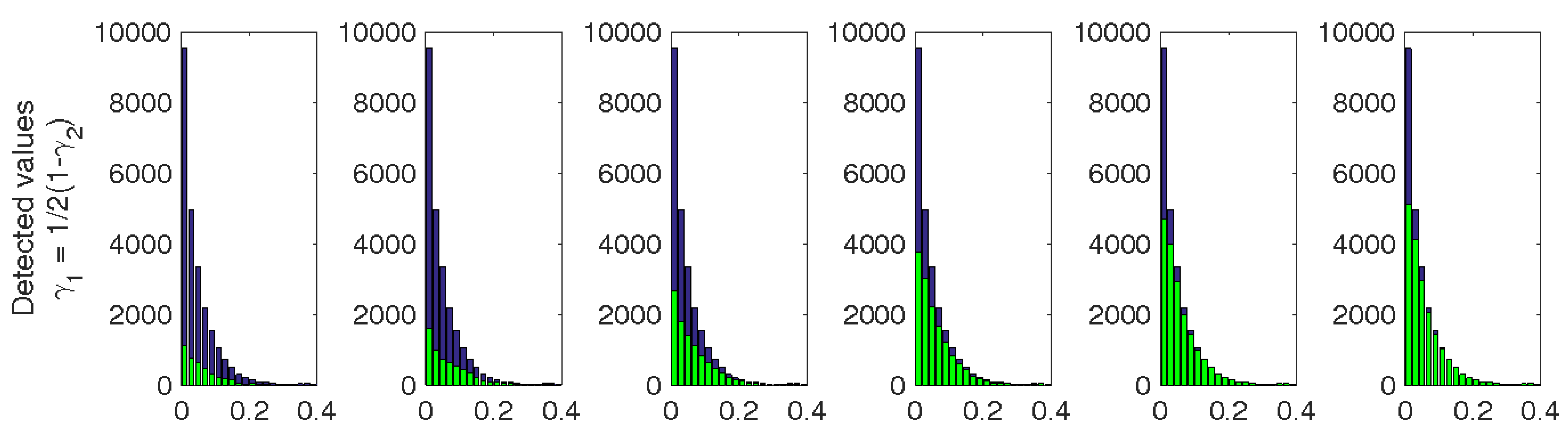}}
\end{center}
\caption{Histograms of the statistically significant values obtained as described in Figure \ref{fig:hist_ic}}
\label{fig:hist_fgc}
\end{figure}

\begin{figure}[H]
\begin{center}
\subfigure{\includegraphics[width=0.8\textwidth]{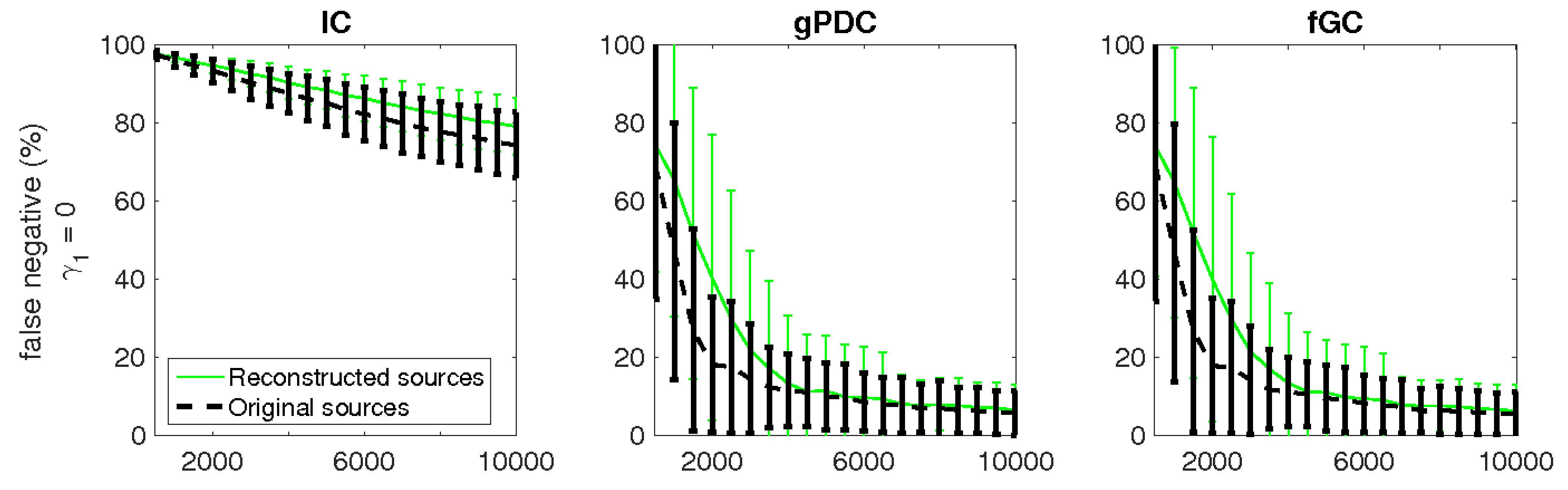}}
\subfigure{\includegraphics[width=0.8\textwidth]{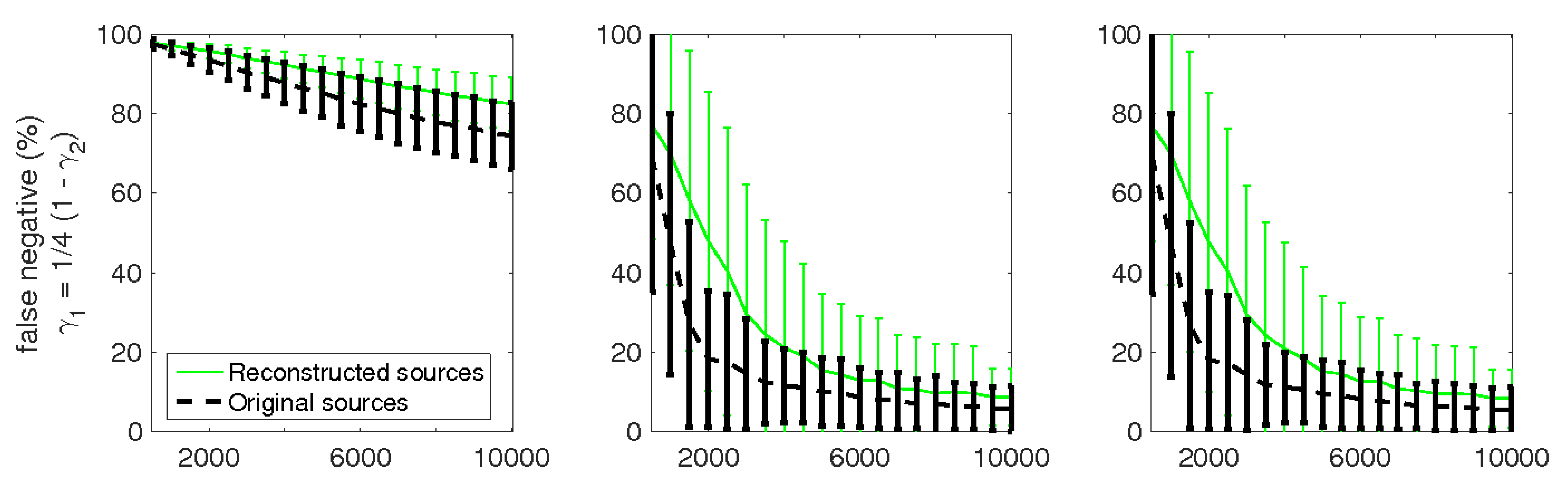}}
\subfigure[Percentage of false negatives]{\includegraphics[width=0.8\textwidth]{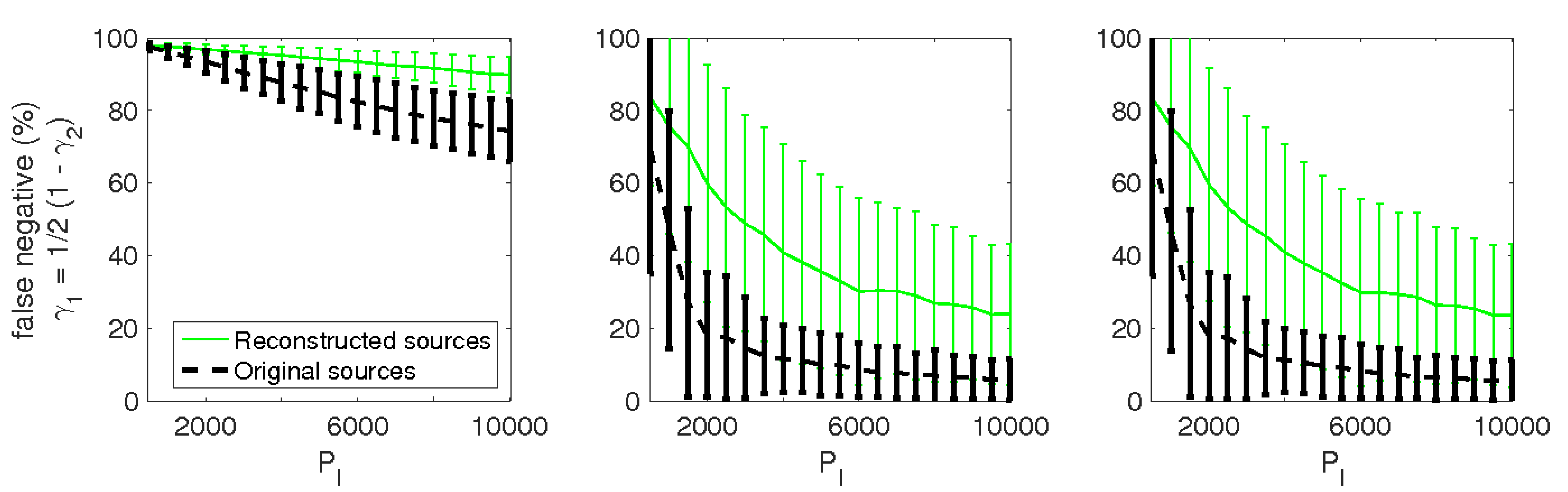}} \\ \vspace{0.1cm}
\subfigure{\includegraphics[width=0.8\textwidth]{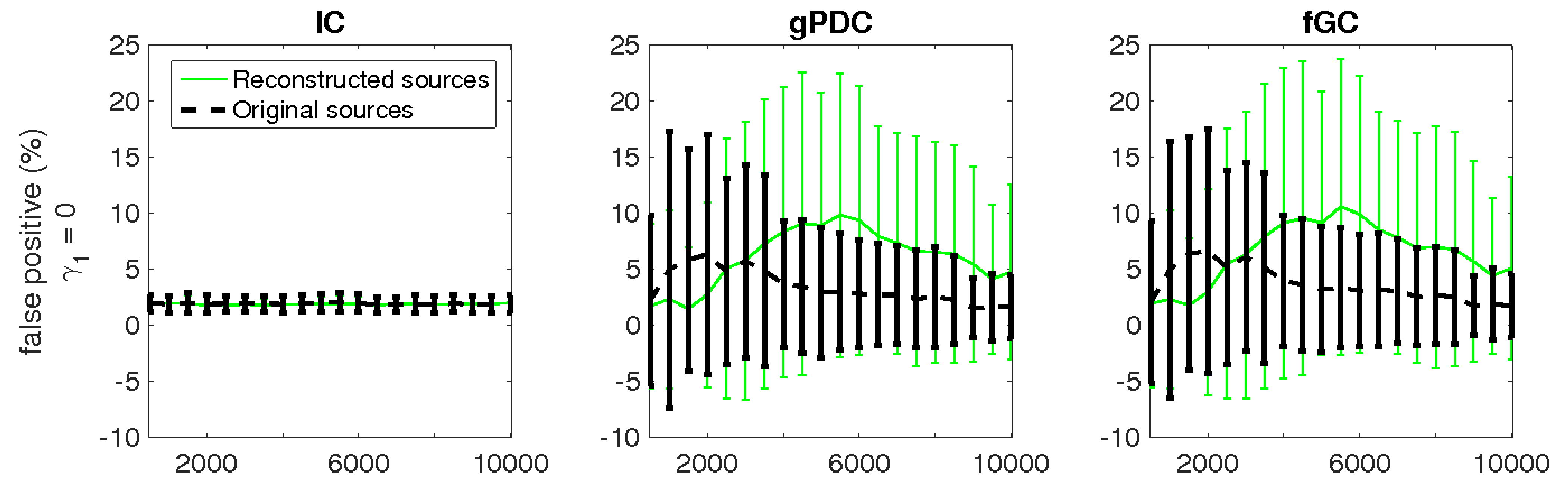}}
\subfigure{\includegraphics[width=0.8\textwidth]{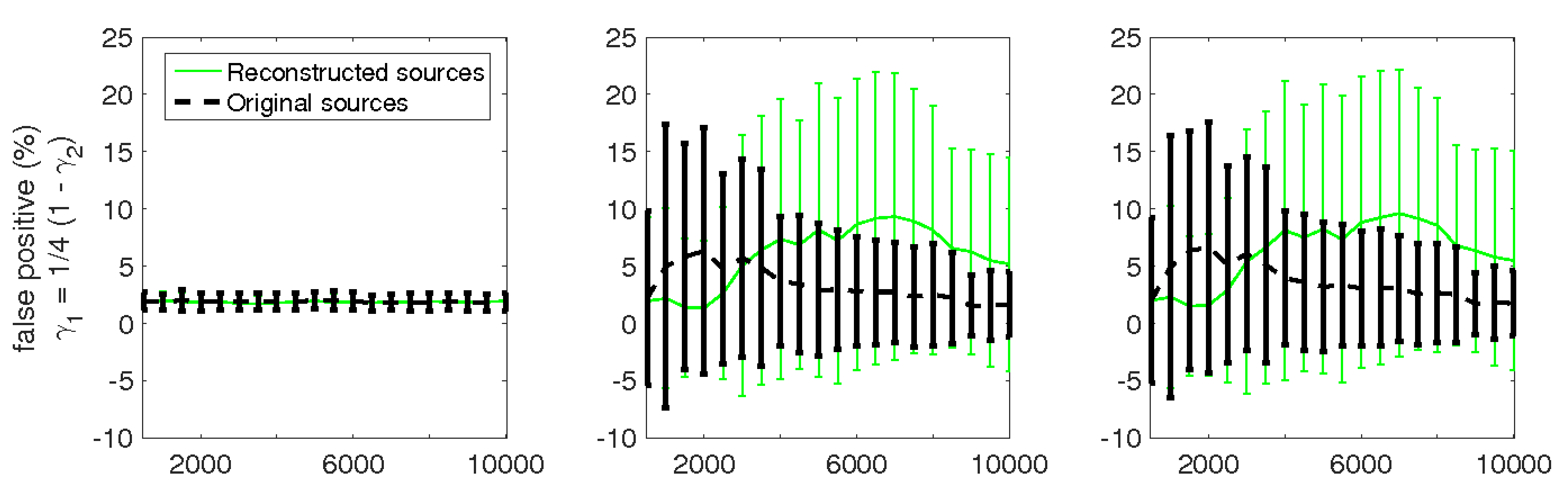}}
\subfigure[Percentage of false positives]{\includegraphics[width=0.8\textwidth]{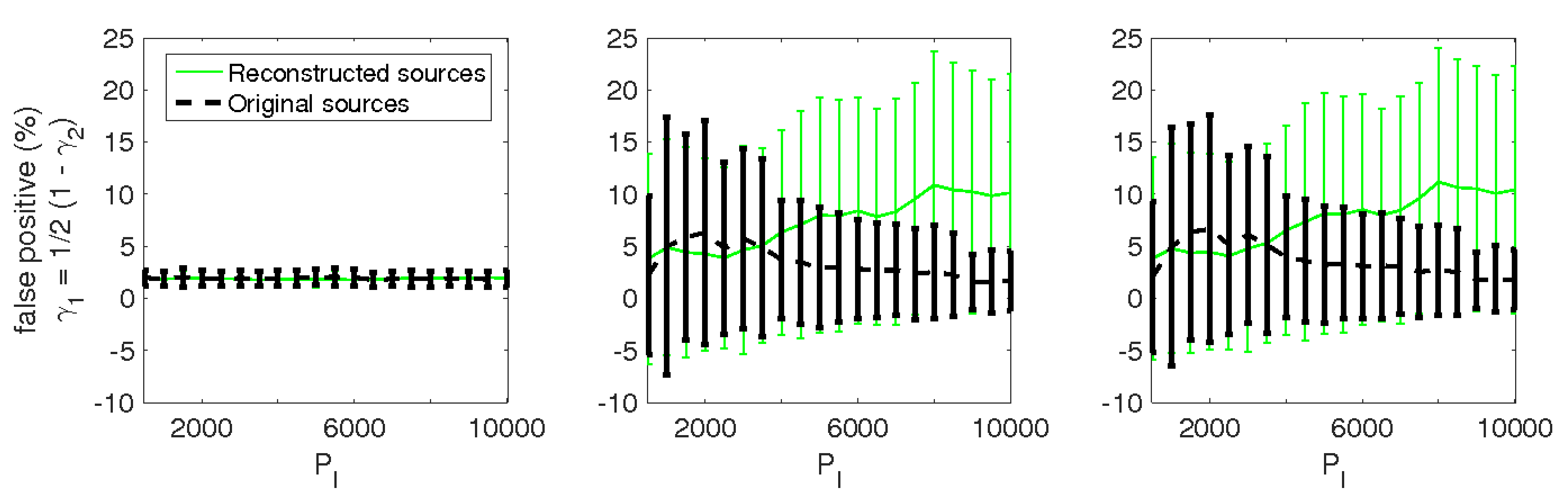}}
\end{center}\caption{First/Last 3 rows: mean and standard deviation over the 100 original datasets of the total percentage of false negatives/positives detected between the pairs of signals $(\tilde{x}_1(t_p), \tilde{x}_2(t_p))$ and $(\tilde{x}_1(t_p), \tilde{x}_3(t_p))$/$(\tilde{x}_1(t_p), \tilde{x}_4(t_p))$, $(\tilde{x}_2(t_p), \tilde{x}_4(t_p))$ and $(\tilde{x}_3(t_p), \tilde{x}_4(t_p))$. In each panel we compare the results obtained from the original source time courses with those from the ROIs activity reconstructed from EEG data with different levels of biological noise.}\label{fig:srec_false_positive}
\end{figure}

\begin{figure}[H]
\begin{center}
\subfigure{\includegraphics[width=\textwidth]{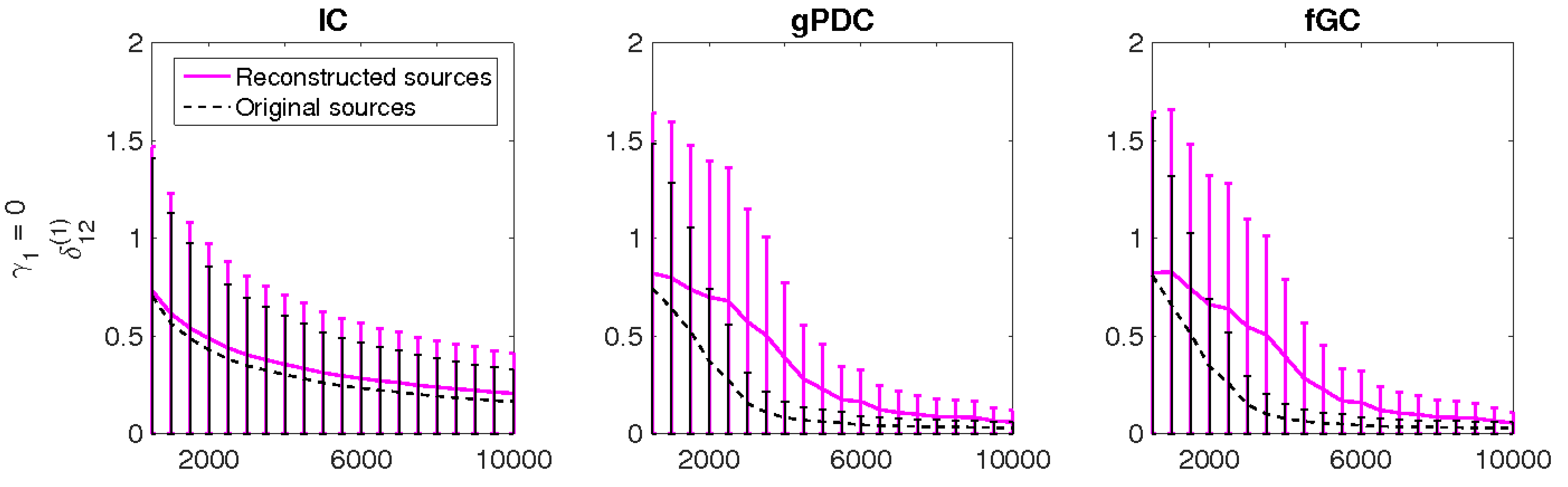}}
\subfigure{\includegraphics[width=\textwidth]{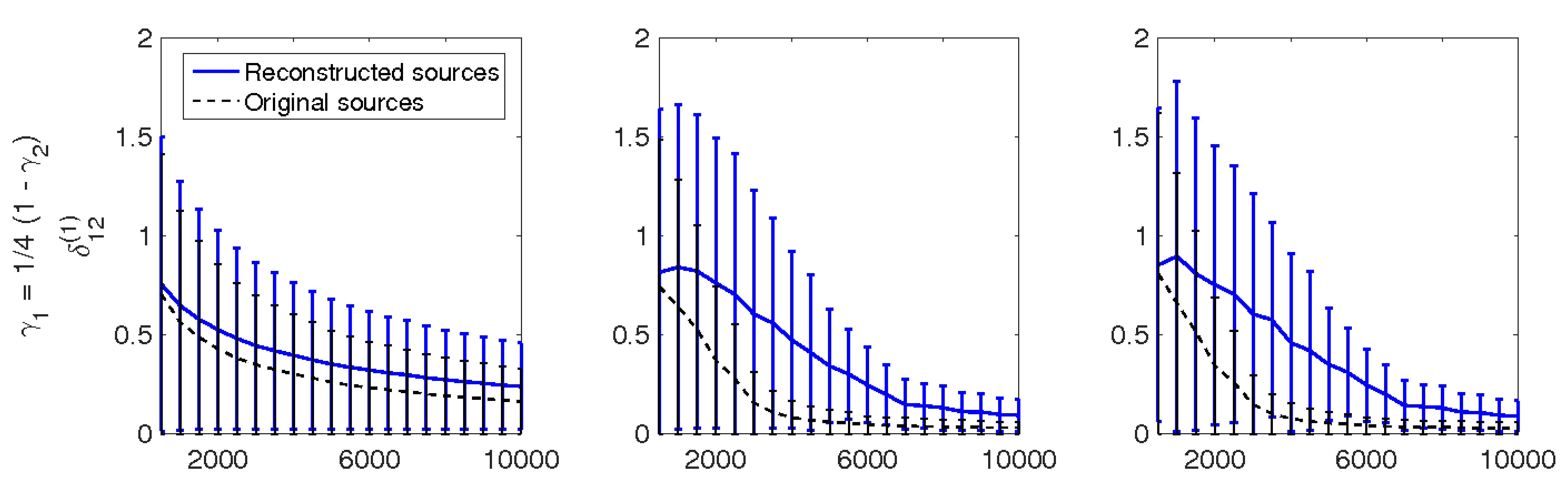}}
\subfigure{\includegraphics[width=\textwidth]{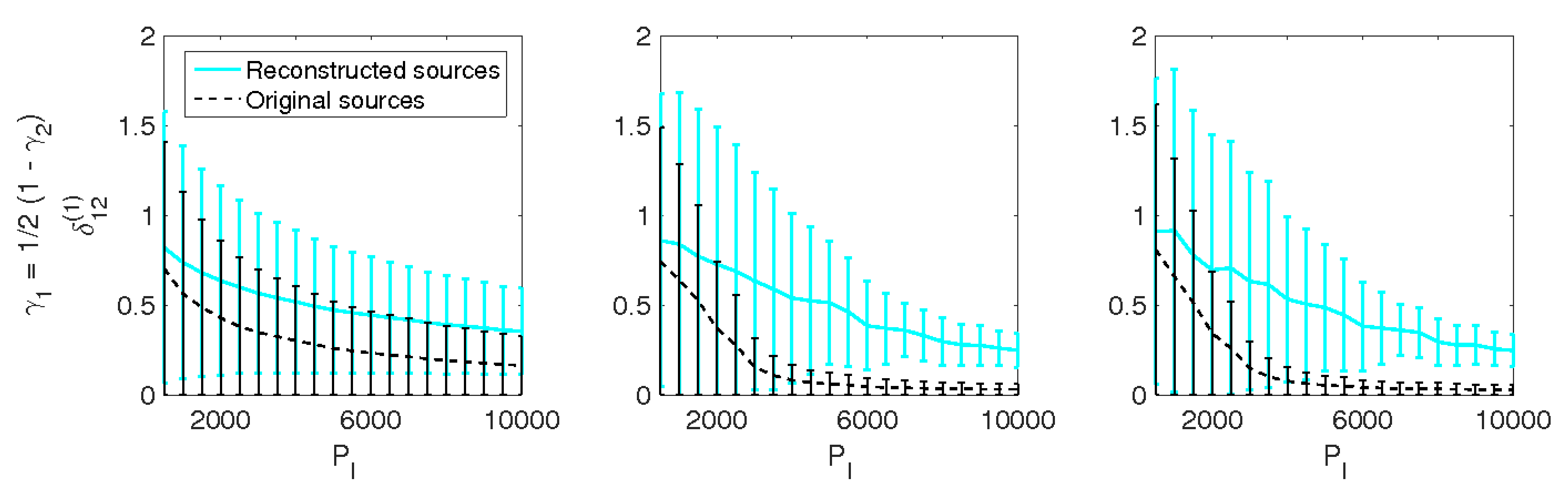}}
\end{center}\caption{Correlation coefficient between the true and the estimated values of the connectivity measures. In each panel we plot the mean and standard deviation over the original datasets of $\delta_{12}$ as a function of the sub--samples length. For each panel the black dotted line represents the correlation coefficient between the true values of the connectivity measure and the empirical values estimated from the true source time courses whereas coloured lines are obtained by estimating the connectivity measured from the reconstructed ROIs activity. Each row concerns a different level of biological noise.}\label{fig:srec_histogram}
\end{figure}

The results of the previous subsection are to be considered as ``optimal'', in the sense that they have been obtained from the original time series.
We now proceed to investigate the effect of the EEG forward/inverse model, using eLORETA, and of biological noise.
In Figure \ref{fig:source_amplitude}, we summarize the analysis pipeline for connectivity analysis from the synthetic EEG data. In the last row of the same figure we exemplify the effect of biological noise on the connectivity estimates, when the connectivity values are estimated from the longest sub--sample. The values of gPDC and fGC are systematically and increasingly under-estimated and over--smoothed: the two bumps of the original profile become less and less visible. Conversely IC seems to be less influenced by the presence of biological noise: for the lower $\gamma_1$ also the estimated IC values are slightly underestimated but still similar to those estimated from the original source time courses (black dotted line). We will quantify this similarity later by making use of the covariance distance $\delta_{ij}$. \\ 
In Figures \ref{fig:hist_ic} -- \ref{fig:hist_fgc} we show the histograms of statistically significant values\footnote{We show only the results obtained when the statistical test is performed using AR surrogate data. The results with PR surrogate data are almost equal.}, grouped according to the true value of the underlying connectivity; different rows correspond to different levels of biological noise. A comparison with the plots in Figure \ref{fig:histogram} suggests that the effect of increasing the noise level is similar to that of shortening the data length. For example, the results with $P=10000$ with measurement noise only (top row, right panel of Figures \ref{fig:hist_ic}--\ref{fig:hist_fgc}) are similar to those from the original time courses with $P=8000$; results for high biological noise (bottom row) with $P=10000$ are similar to those for measurement noise only (top row) with $P=4000$. \\
In Figure \ref{fig:srec_false_positive}, we show the false negative and false positive rate for different values of the SNR. For the sake of comparison, we also superimpose the results from the original time courses. Expectedly, the false negative rate increases with increasing biological noise for all the connectivity measures. More interestingly, the plots obtained from the reconstructed sources show a higher number of false positives at increasing data length only for gPDC and fGC.
These results seem to confirm that the effect of the biological noise we simulated is not qualitatively different from the effect of having less data.\\
Finally, in Figure \ref{fig:srec_histogram} we plot the mean and standard deviation over test dataset of the correlation distance $\delta_{12}$ defined in equation (\ref{eq:correlation_distance}), for the pair of correlated sources $\left( x_1(t_p), x_2(t_p) \right)$. We recall that this correlation distance does not depend on the significance of the estimated values; instead, it is an attempt at quantifying how much of the true connectivity pattern, as a function of frequency, is retained in the estimated connectivity pattern.
Here we show the values of the correlation distance for connectivity estimated from the original time courses (black dotted line) and from the estimated time courses (coloured lines). The plots for IC are easier to interpret. The effect of increasing the data length is a monotonic decrease of the correlation distance; noise, on the other hand, monotonically increases such distance. For gPDC and fGC, the connectivity estimates from the original time courses become quickly fairly good, as the data length increases. However, the EEG forward/inverse models and biological noise disrupt such good behaviour more dramatically than for IC.
The plots suggest that gPDC and fGC are still preferable for longer data lengths, where they have smaller variance and smaller/comparable correlation distance. The case of high biological noise, however, provides rather similar performances for all three metrics.

%
%

%% file: discussion.tex
\section{Discussion}

In the present work we studied the impact of the length of the time series on the estimates of connectivity from EEG data, using three frequency--domain connectivity measures, IC, gPDC and fGC, and numerical simulations.
We first studied the ideal case, in which the source time courses are known exactly, to provide a reference value and isolate the effects due solely to the finiteness of the data. Then we went on to consider the case where connectivity is computed from source time courses that are, in turn, estimated from EEG time series, possibly affected by biological noise. We quantified the impact of the various elements in multiple ways: by assessing the significance of the estimated connectivity values, using surrogate data; by computing the false negative and false positive rates; by computing a correlation distance with the theoretical connectivity value, which is known exactly.\\

\noindent
As mentioned in the introduction, there are few studies aimed at quantifying the impact of the data length on the quality of the connectivity estimates.
However, to the best of our knowledge, this is the first study that takes into account the effect of the EEG forward/inverse model and of biological noise, and compares connectivity from the original time courses with connectivity from the estimated time courses. \\

\noindent
Our numerical results from the exact time courses confirm most expectations concerning qualities and drawbacks of the connectivity metrics under investigation.
Specifically, all the simulations confirm that IC is much more ``conservative'' than gPDC and fGC, meaning that the estimated IC values are less likely to pass the threshold of statistical significance, compared to the estimated gPDC and fGC values. In fact, the empirical values of IC tend to pass the statistical test only in correspondence of the greater theoretical values, providing a systematically higher number of false negatives. On the other hand, estimated gPDC and fGC values pass the threshold very often and tend to be more affected by false positives. In general, it can be said that these two model--based measures tend to provide more accurate values than IC when the MVAR model is correctly estimated, but are more prone to errors when it is not. This is particularly manifest in the case of a common input: if the common input is correctly addressed in the model, gPDC and fGC provide better results than IC, which is a bi--variate measure; if it is not, results from gPDC and fGC are less reliable. This issue might be particularly relevant in MEG studies, where radial sources might not show up in the sensor level data. \\

\noindent
The numerical results from the estimated time courses clearly indicate that the effects of the EEG forward/inverse plus measurement noise, and of biological noise are not negligible.
Even in the absence of biological noise, the results obtained from the estimated time courses are worse than those obtained from the exact time courses; in particular, they appear similar to those obtained from the exact time courses but with shorter data. This is reasonable because measurement noise lowers the signal-to-noise ratio without disrupting connectivity. Additionally, the use of ROIs (centred in the correct dipole locations) in place of pointlike sources only partially compensates for the lowering of the signal-to-noise ratio in the reconstructed source induced by the spatial dispersion of the forward/inverse procedure.\\
The effect of biological noise does not seem to be qualitatively different, i.e. mostly has the same effect of having shorter time series, but is quantitatively relevant, yielding substantial increased false negative and false positive rates; the only exception here appears to be the false positive rate of IC, which remains systematically low.\\
Our results seem to indicate that explicitly modelling the forward/inverse procedure is key for a quantitative assessment of the reliability of connectivity measures. We speculate that this is particularly relevant when comparing data--driven and model--based measures. Indeed, in our simulations the results obtained from the estimated sources and those obtained from the original sources appear to be more diverse for gPDC and fGC than for IC. Comparative studies that do not explicitly model the forward/inverse procedure might thus be prone to misleading conclusions.\\

\noindent
Our results are of course limited by the type of simulations we performed, and might still be optimistic. For instance, biological noise is simulated
as the activity of ten independent sources, and the regularization parameter in eLORETA has been chosen based on the knowledge of the noise variance.
Even more so, our results show that the impact of the EEG forward/inverse and of biological noise should be taken into account when one wants to quantitatively
assess the sensitivity of the connectivity metrics to the data length; if not, results are overly optimistic.


%% file: conclusions.tex
\section{Conclusions}

The great advantage of EEG and MEG in the study of brain connectivity resides mainly in their temporal resolution, that allows, in principle, monitoring
of dynamic changes in the connectivity patterns. However, this requires connectivity to be estimated from short time windows.
This study confirms that such estimation is possible, but with some caution, because using less data makes the results more prone to errors. 
In addition, some contributions to error, like neglecting a common input in the estimate of MVAR based connectivity metrics, in fact increase with increasing data length.
While none of these empirical results is particularly surprising from a theoretical perspective, their consequences should be taken into account when studying dynamic connectivity from EEG and MEG time series. Finally, our numerical results suggest that an analytic characterization of the impact of the various error causes on the quality of the connectivity estimates would prove useful. Future studies will be devoted to address this point.